\documentclass[preprint,12pt,authoryear]{elsarticle}

\usepackage{float}
\usepackage{amssymb}
\usepackage{textgreek}
\usepackage{upgreek}
\usepackage{url}
\usepackage[utf8x]{inputenc}
\usepackage[T1]{fontenc}
\usepackage{amsmath}
\usepackage{amsfonts}
\usepackage{amssymb}
\usepackage{graphicx}
\graphicspath{{images/}}
\usepackage{fancyhdr} 
\usepackage{vmargin}
\usepackage{verbatim}
\usepackage{lineno}
\usepackage{hyperref}
\usepackage{comment}
\usepackage{subcaption}
\usepackage{multirow}

\usepackage{todonotes}

\reversemarginpar

\begin{document}

\begin{frontmatter}

\title{Toward Ionization Cluster Size Measurements with a Compact Nanodosimeter}
\tnotetext[t1]{Prepared for submission to RADMEAS}

\author[label1,label2,label3]{Victor Merza\corref{cor1}}
\ead{victor.merza@tecnico.ulisboa.pt}
\cortext[cor1]{Corresponding author}
            
\author[label4,label5]{Aleksandr Bancer}

\author[label3]{Vladimir Bashkirov}
    
\author[label1,label2]{Ana Belchior}

\author[label6]{Beata Brzozowska}

\author[label7,label8]{Piotr Gasik}

\author[label5]{Jaroslaw Grzyb}

\author[label1,label2]{Khaled Katmeh}

\author[label4,label5]{Marcin Pietrzak}

\author[label4]{Antoni Ruciński}

\author[label3]{Reinhard Schulte}

\affiliation[label1]{organization={Centro de Ciências e Tecnologias Nucleares, Instituto Superior Técnico, Universidade de Lisboa},
            addressline={Estrada Nacional 10 (km 139,7)}, 
            city={Bobadela LRS},
            postcode={2695-066}, 
            country={Portugal}}
\affiliation[label2]{organization={Departamento de Física, Instituto Superior Técnico, Universidade de Lisboa},
            addressline={Av. Rovisco Pais 1}, 
            city={Lisboa},
            postcode={1049-001},
            country={Portugal}}
\affiliation[label3]{organization={Department of Basic Science, Division of Biomedical Engineering Sciences, Loma Linda University},
            addressline={201 Mortensen Hall, 11085 Campus St}, 
            city={Loma Linda},
            postcode={92350}, 
            state={California},
            country={United States of America}}
\affiliation[label4]{organization={Cyclotron Centre Bronowice, Institute of Nuclear Physics Polish Academy of Sciences},
            addressline={Radzikowskiego 152}, 
            city={Kraków},
            postcode={31-342}, 
            country={Poland}}
\affiliation[label5]{organization={Radiological Metrology and Biomedical Physics Division, National Centre for Nuclear Research},
            addressline={ul. Andrzeja Sołtana 7}, 
            city={Otwock Świerk},
            postcode={05-400},
            country={Poland}}
\affiliation[label6]{organization={Biomedical Physics Division, Institute of Experimental Physics, Faculty of Physics, University of Warsaw},
            addressline={5 Pasteur Street}, 
            city={Warsaw},
            postcode={02-093}, 
            country={Poland}}
\affiliation[label7]{organization={GSI Helmholtzzentrum f\"ur Schwerionenforschung GmbH (GSI)},
            addressline={Planckstr. 1}, 
            city={Darmstadt},
            postcode={64291}, 
            country={Germany}}
\affiliation[label8]{organization={Facility for Antiproton and Ion Research in Europe GmbH (FAIR)},
            addressline={ Planckstr. 1}, 
            city={Darmstadt},
            postcode={64291},
            country={Germany}}

\begin{abstract}
Nanodosimetry aims to provide measurable quantities related to the nanoscopic particle track structure, which determines the biological effectiveness of radiation.
While simulated nanodosimetry has already demonstrated its potential for radiation treatment planning, the experimental realization of practical nanodosimetric detectors is still in its early stages.

In this work, a nanodosimetric prototype operated with low-pressure gas was developed to count ionizations in a nanometer-equivalent sensitive gas volume.
Its performance was evaluated experimentally with alpha beams from an $^{241}$Am source in 1~mbar propane gas.
The results support the further development of this compact nanodosimeter class, with potential applications in particle therapy, radiation protection, and space radiation dosimetry.
\end{abstract}


\end{frontmatter}

\section{Introduction}
Nanodosimetry aims to predict the frequency and complexity of DNA damage based on the track structure of ionizing particles, which exhibits a nanometric spatial distribution of excitations and ionizations in tissue \citep{rucinski,conte2023,Faddegon}.
Ionizations are considered the primary physics contributors to radiobiological effectiveness, either by producing diffusing radicals that cause indirect DNA damage or by directly ionizing the DNA molecule.
When ionizations occur in a clustered pattern, the probability of inducing irreparable complex lesions becomes significant for more than four ionizations within or near the DNA molecule.
The ionization cluster size, i.e., the number of ionizations produced by a primary particle and its secondaries in a nanometric target volume, such as a DNA segment of about 10–20 base pairs with a length of about 3.4 nm to 7.8 nm, is proportional to the resulting DNA lesion cluster size.
Nanodosimetric quantities like the ionization cluster size distribution (ICSD), describing the spatial distribution of ionizations, are measurable.
Thus, by characterizing the nanometric pattern of ionization clustering, nanodosimetry provides an experimentally accessible alternative descriptor of radiation quality \citep{conte2017,Ortiz2025}.

State-of-the-art technology does not allow the direct measurement of nanodosimetric quantities in tissue or tissue-equivalent materials such as water.
However, nanodosimetric quantities can be measured in low-pressure gases and thus in nanometer-equivalent sensitive volumes (SVs).
This nanometer-scale equivalence of millimeter-sized SVs at typical pressures of 1 mbar to 3 mbar is achieved by applying density scaling theorems \citep{grosswendt1,grosswendt2}, accounting for radiation quality, gas type, and mass per area.
The few fully operational nanodosimeters working at low gas pressure developed to date are inherently complex systems.
These include the ion counter detector \citep{garty2002,hilgers2015,hilgers2019,hilgers2022}, the LNL Startrack Counter \citep{denardo2002}, and the Jet Counter \citep{pszona2000,bantsar,pietrzak2018,bancer2020}.
Recent efforts have focused on developing more compact track-structure imaging detectors with nanometer-equivalent resolution \citep{bashkirov20091,bashkirov20092,casiraghi2014,casiraghi2015} and prototypes focusing on investigating the basic working principle \citep{vasi2016,merza2026} or measuring gas transport parameters \citep{kempf20252}.
These detector prototypes combine concepts from thick gas electron multipliers (THGEMs) and resistive plate chambers (RPCs), but operated at low pressure and in reverse polarity.
Here, positive ions produced in the SV are guided into sub-millimeter to few-millimeter-diameter holes in dielectric plates with thicknesses of a few millimeters to about 10~mm.
Inside the hole, due to the application of a strong electric field, the collected ions initiate a charge avalanche, which induces a measurable signal.
Due to the low diffusion of the collected ions, a nanometer-equivalent spatial resolution of the ion starting positions and thus a detailed reconstruction of the underlying ionization pattern can be achieved.

Up to now, no compact nanodosimetric detector has been used to measure the ICSD in a single SV representing a short segment of the DNA molecule.
With the techniques applied so far, this appears to be only feasible with grouped multi-hole amplification structures that employ independent holes that create separate signals for separate ions collected from the same nanometer-equivalent SV \citep{merza2025,merza2026}.
Furthermore, the ion detection efficiency of future compact nanodosimeter prototypes must be improved before they can be reliably applied for the measurement of nanodosimetric quantities.

The goal of the measurements presented in this paper was to demonstrate a step toward the development of miniaturized nanodosimeters based on grouped multi-hole amplification structures capable of measuring the ICSD.
Such nanodosimeters may ultimately be applied in radiation therapy.

\section{Materials and Methods}
\subsection{Detector Working Principle}
The presented detector is a compact nanodosimeter featuring multiple grouped holes drilled into dielectric plates, in which the signal formation occurs.
It is designed to determine the number of ionizations produced in the low-pressure gas above the dielectric plates by incident radiation by extracting and detecting the positive ions generated in the gas.

The detector design is conceptually inspired by THGEM and RPC technologies, while operating at low gas pressure and in reverse polarity compared to conventional implementations.  
Related detector systems based on charge multiplication initiated by positive ions include the TIDe~\citep{bashkirov20091,bashkirov20092}, FIRE~\citep{FIRE}, and FIRE-V2~\citep{kempf20252} nanodosimeters, as well as other prototypes \citep{vasi2016,merza2026}.  

\begin{figure}[htbp]
    \centering
    \includegraphics[width=0.8\linewidth]{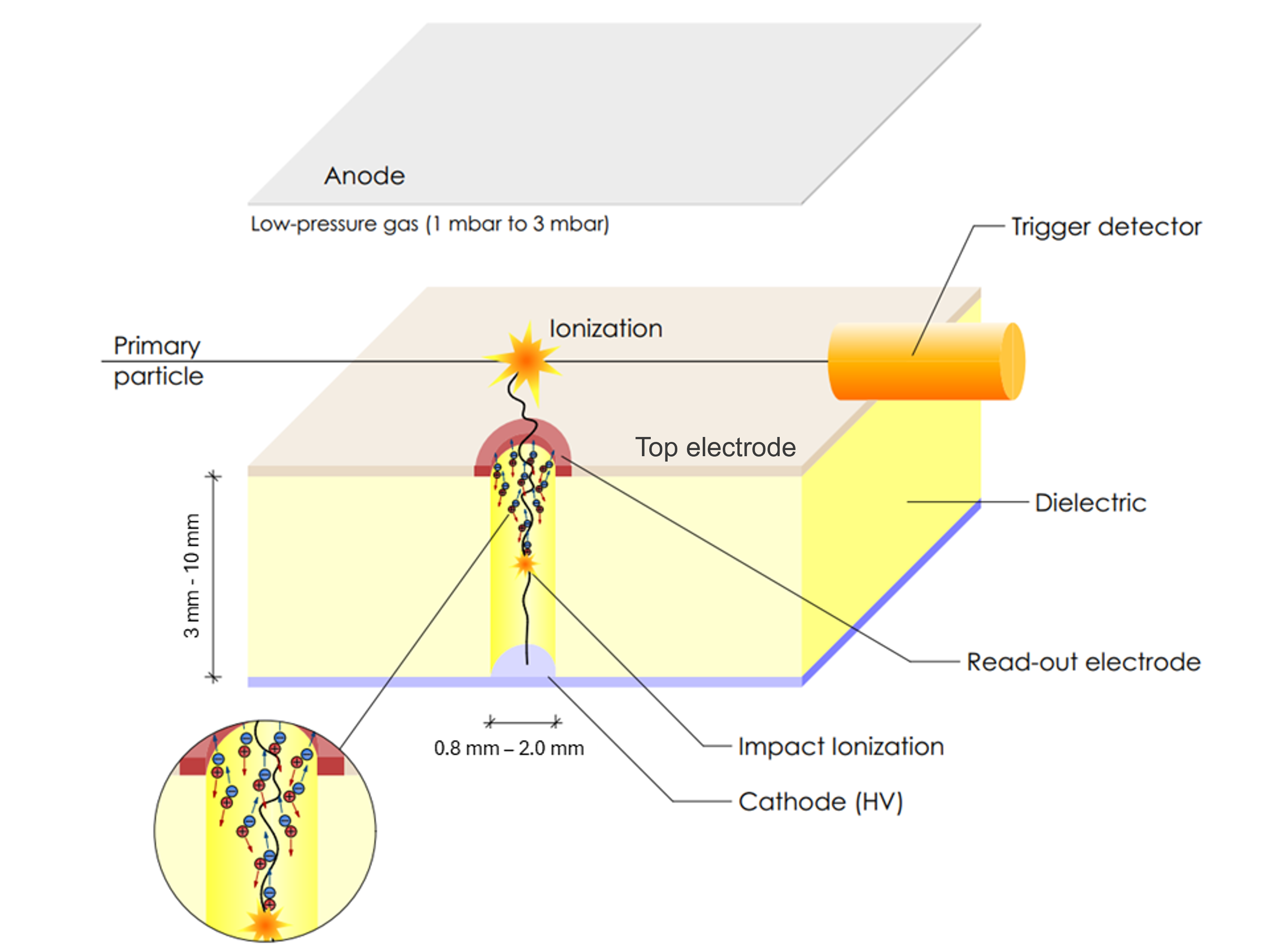}
    \caption{Illustration of the proposed working principle of contemporary compact nanodosimeters based on ion multiplication in low-pressure gas. While the mechanism initiating the avalanche is not yet fully understood, the figure illustrates the ion-impact ionization process proposed in \citep{bashkirov20091}. Ion-induced secondary electron emission from the cathode (not shown) may also contribute to, and possibly dominate, signal formation \citep{merza2026}.}
    \label{fig_working_principle}
\end{figure}

The underlying working principle is illustrated in Fig.~\ref{fig_working_principle}, which shows a single-hole realization for clarity.
The same principle applies to multi-hole detectors.
A beam of primary particles enters the detector, filled with a low-pressure gas, and is registered by a trigger detector.
An electric drift field is created between the anode plate and the grounded top electrode attached to a dielectric plate.
The dielectric plate contains one or multiple holes surrounded by a readout electrode.
A resistive glass cathode mounted on the bottom side of the dielectric plate generates a strong electric field within the holes.
Positive ions produced by the primary particles drift towards the hole(s), where they are collected.
The low ion diffusion enables high spatial resolution of the initial ion positions, thereby allowing a detailed reconstruction of the underlying nanometer-equivalent ionization pattern.
Within a hole, the ions may initiate a charge avalanche, which induces a measurable signal on the readout electrode.
The avalanche is initiated by the release of an electron, either through ion-impact ionization of a gas molecule or through ion-induced secondary electron emission (IISEE) from the cathode.
The relative contribution of these two signal formation mechanisms remains to be further investigated \citep{merza2026}.

\subsection{Experimental setup}
The experimental setup utilizes the same low-pressure chamber and most components of the data acquisition system as described in \citep{merza2026}.

\begin{figure}[htbp]
    \centering
    \includegraphics[width=0.6\linewidth]{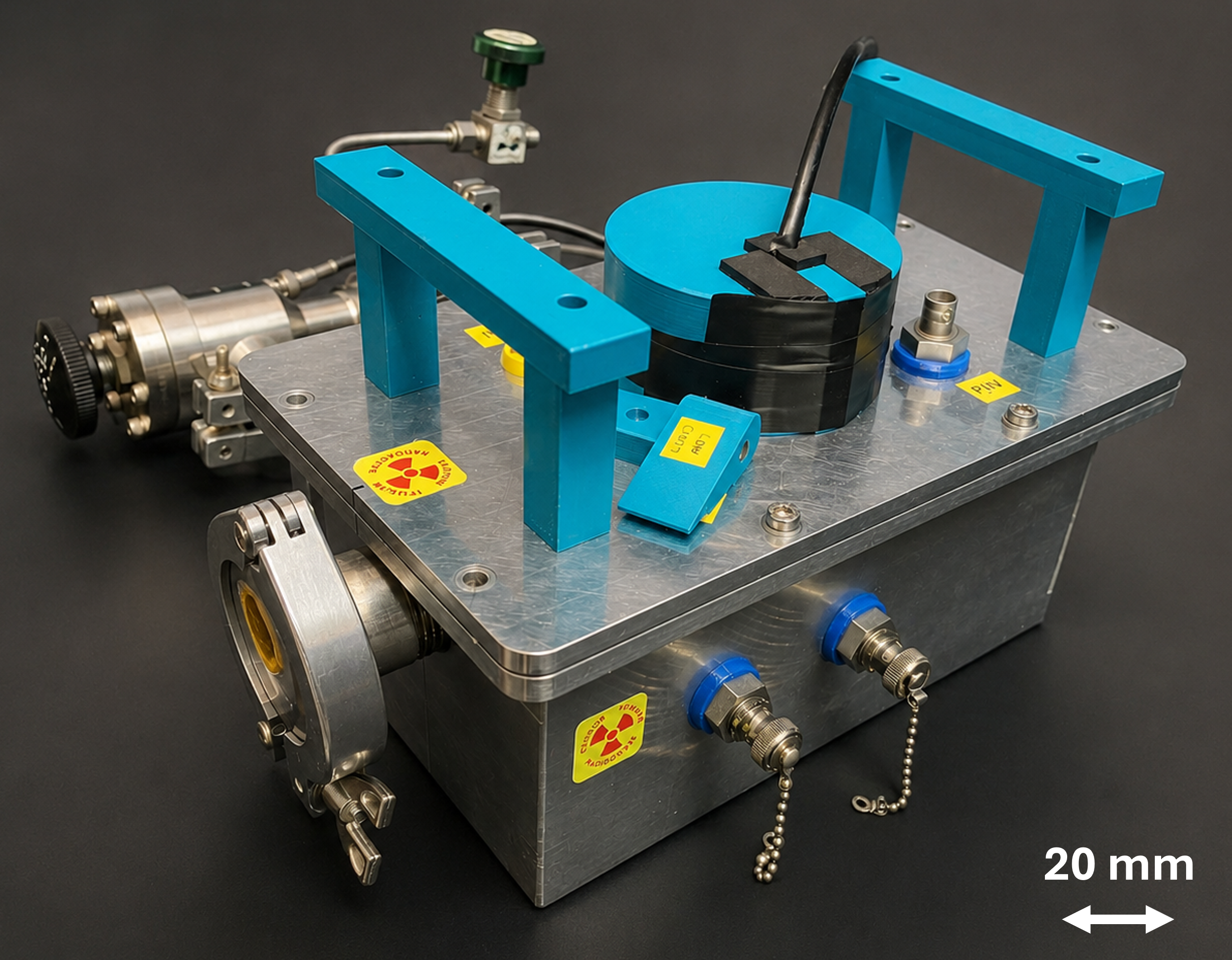}
    \caption{Low-pressure chamber used for the experimental setup.}
    \label{fig_low_pressure_chamber}
\end{figure}

\begin{figure}[htbp]
    \centering
    \includegraphics[width=0.8\linewidth]{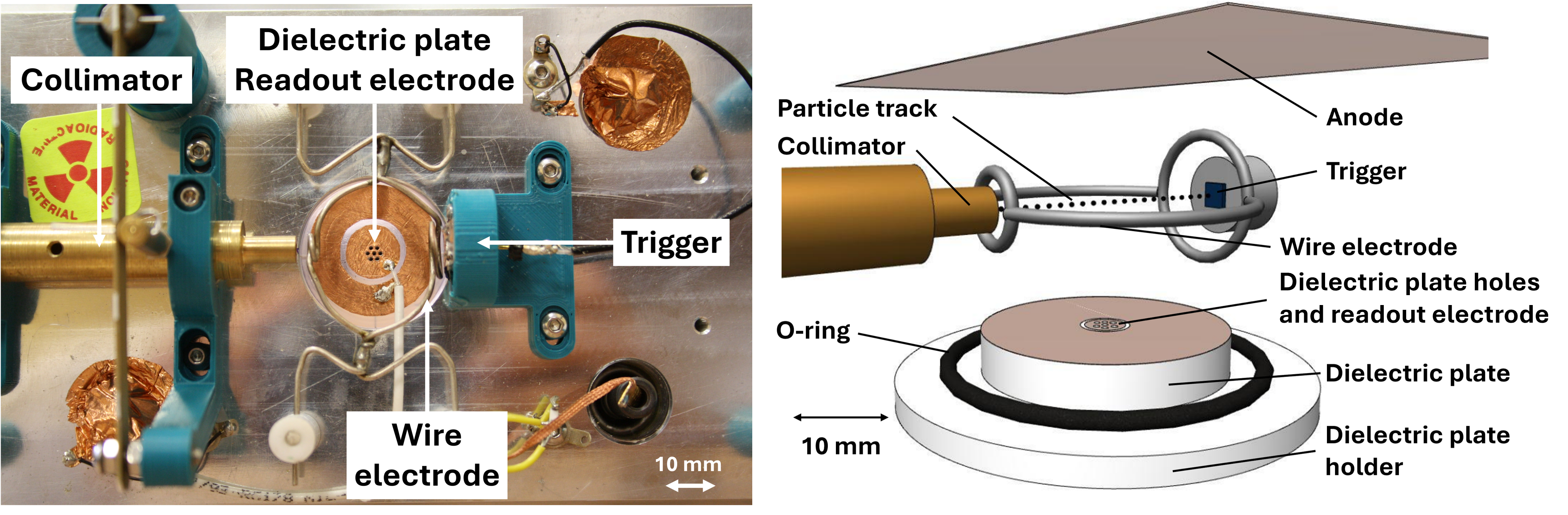}
    \caption{Top view of the 
    detector interior, equipped with dielectric plate D-3, attached to the underside of the low-pressure chamber lid.}
    \label{fig_detector_D3}
\end{figure}

\begin{figure}[htbp]
    \centering
    \includegraphics[width=0.8\linewidth]{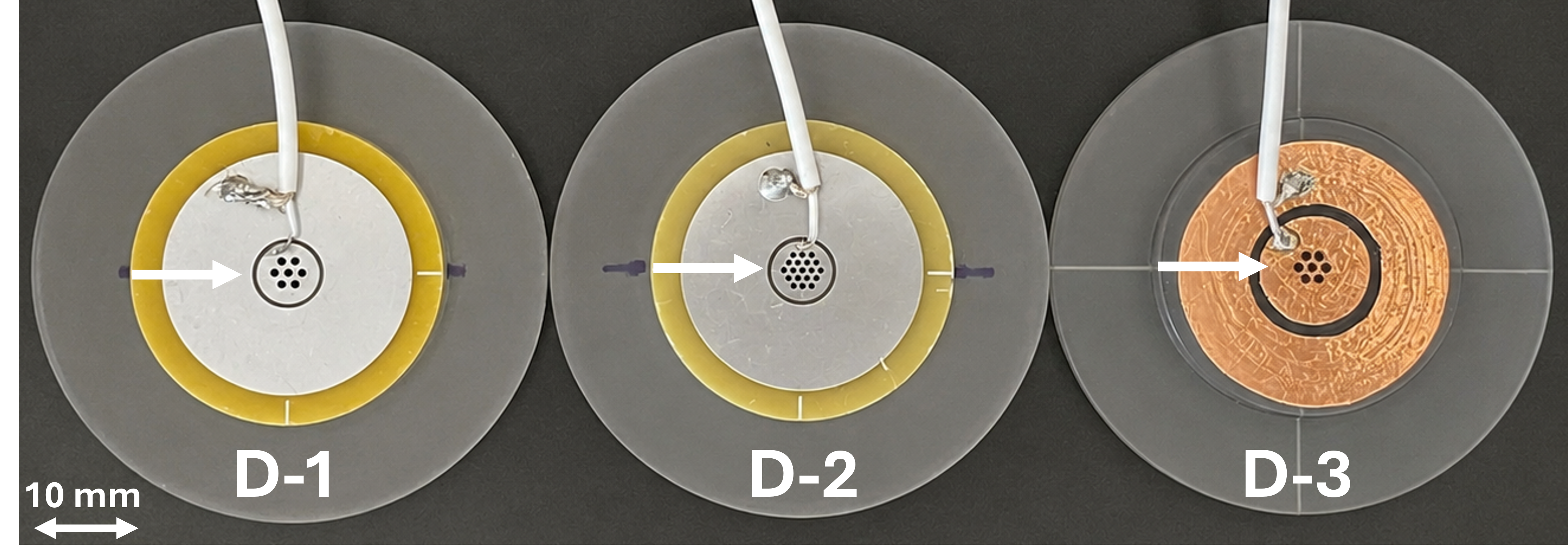}
    \caption{Dielectric plates D-1, D-2, and D-3, attached to their respective acrylic holders. The white arrows indicate the beam direction.}
    \label{fig_plates}
\end{figure}

\paragraph{Low-pressure chamber and dielectric plates}\label{meth_dielectric_plates}
Fig.~\ref{fig_low_pressure_chamber} shows the low-pressure chamber of the detector setup used in this work.
Fig.~\ref{fig_detector_D3} shows the detector interior mounted to the underside of the lid of the low-pressure chamber.
Three distinct versions of dielectric plates featuring multiple grouped holes were manufactured and are shown in Fig.~\ref{fig_plates}.
They differ in dielectric material, plate thickness, hole diameter, and hole multiplicity (see Tab.~\ref{tab_dielectric_plates}).
Two dielectric plates (D-1 and D-2) employ 3.175~mm thick plates fabricated from FR-4 150~TG, whereas the third dielectric plate (D-3) uses a 5~mm thick acrylic plate.
All dielectric plates feature sub-millimeter holes arranged in a honeycomb-like geometry.
D-1 and D-2 are glued to circular acrylic holders to fit into an opening in the low-pressure chamber lid, while D-3 was manufactured as a one-piece unit, including the holder.

\begin{table}[htbp]
    \centering
    \begin{tabular}{l|l|l|l}
         Dielectric plate & D-1 & D-2 & D-3 \\ \hline
         Material & FR-4 150 TG & FR-4 150 TG & Acrylic \\
         Thickness & 3.175~mm & 3.175~mm & 5 mm \\
         Number of holes & 7 & 19 & 7 \\
         Hole diameter & 0.8~mm & 0.6~mm & 0.8~mm \\
         Hole pitch & 1.3~mm & 0.9~mm & 1.3~mm \\
    \end{tabular}
    \caption{Overview of the characteristics of the dielectric plates used in this work.}
    \label{tab_dielectric_plates}
\end{table}

D-1 consists of a plate with 7 holes of 0.8~mm diameter and a pitch of 1.3~mm, while D-2 incorporates 19 holes of 0.6~mm diameter with a pitch of 0.9~mm.
D-3 contains 7 holes of 0.8~mm diameter arranged with a pitch of 1.3~mm.
The hole patterns of D-1 and D-3 lie within a circle with a diameter of approximately 3.4~mm, while the corresponding diameter for D-2 is approximately 4.2~mm.
Applying the density scaling procedure of Grosswendt et al.~\citep{grosswendt1,grosswendt2}, at 1~mbar propane, these diameters correspond to approximately 8.4~nm (D-1 and D-3) and 10.4~nm (D-2) when scaled to liquid water with unit density for the incident alpha particles with a mean energy of about 4.6~MeV investigated in this work.
The actual SV size and shape, which also depends on the applied cathode and anode potentials, deviate from the geometrical hole pattern, as it is defined by ion transport and collection rather than a sharp boundary.
Ions generated outside may still be collected within the SV, while ions generated inside may be lost due to diffusion and field-driven transport.
A dedicated SV mapping, as presented in \citep{merza2025} for a single-hole plate of a compact nanodosimeter, was not carried out in the present work, as it was beyond its scope.
For D-1 and D-2, the inner readout electrodes, with a diameter of 6~mm, and the top electrodes, with a diameter of 24~mm, were made of tin–lead.
In D-3, both electrodes were made of copper foil affixed to the dielectric plate, with the slightly larger readout electrode having a diameter of approximately 10~mm.

The corresponding $pd$ (gas pressure $p$ of 1~mbar propane times dielectric plate thickness $d$) values of these amplification structures are approximately $0.32\,\rm Pa\,m$ for D-1 and D-2, and $0.5\,\rm Pa\,m$ for D-3.
Accordingly, the detector operates below the Paschen minimum, with the minimum for propane reported at approximately $0.67\,\rm Pa\,m$ \citep{heylen1975}.

\begin{figure}[htbp]
    \centering
    \includegraphics[width=0.6\linewidth]{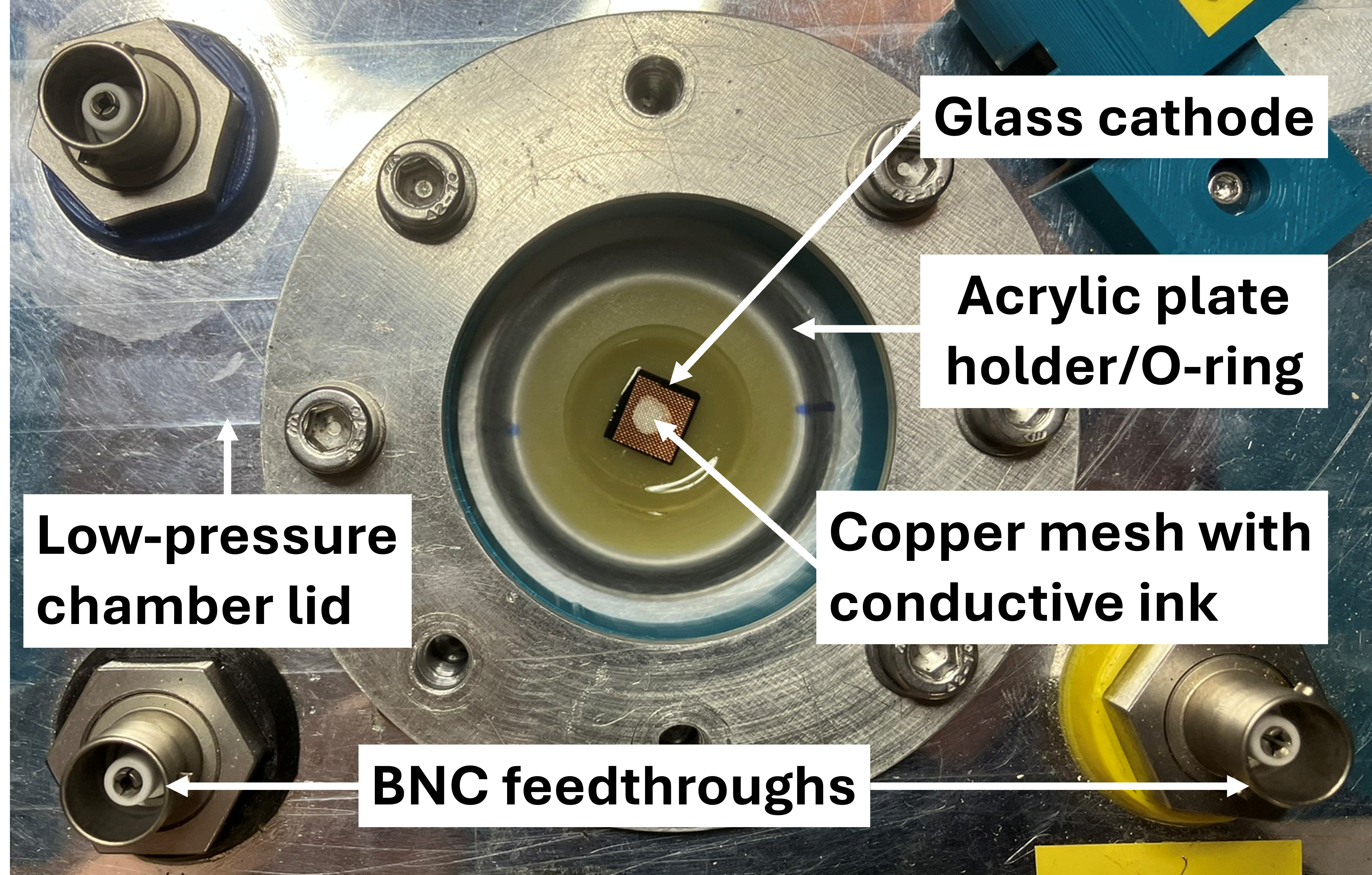}
    \caption{Top view of the opening in the detector lid, into which the dielectric plate D-1, equipped with a glass cathode, is inserted. The copper mesh and conductive ink were used to ensure a reliable electrical contact with the HV supply.}
    \label{fig_plate_in_lid}
\end{figure}

\paragraph{Cathode, electrodes, and voltage supply}\label{meth_cathode_electrodes_voltage}
A 0.7~mm thick low-resistivity glass cathode (6~mm~$\times$~6~mm) was mounted on the bottom of the dielectric plate (Fig.~\ref{fig_plate_in_lid}).
The bulk resistivity of the glass is on the order of $10^{10}~\Omega$~cm \citep{wang2019,wang2010}.
The cathode was placed in direct contact with the dielectric plate and sealed using a non-conductive silicone sealant to ensure mechanical stability and gas tightness.
This silicone sealant was applied along the cathode–dielectric plate contact edge from the air side.

A rectangular copper anode plate with dimensions of 80~mm$~\times$~40~mm was positioned 20~mm above the dielectric plate to establish the drift field.
To obtain a nearly uniform electric field in the drift region, a circular silver-clad copper wire electrode surrounded the active volume and incorporated ring-shaped beam apertures to minimize field distortions from adjacent detector components (Fig.~\ref{fig_detector_D3}).
The central axis of the primary particle beam traversed the drift region at a height of 10~mm and passed directly above the center of the hole array in the dielectric plate.

The cathode potential was varied between $-600$~V and $-1000$~V for the different dielectric plates and for a propane gas pressure of 1~mbar.
The anode was maintained at 15~V, while the wire electrode was set to 7.5~V.
The dielectric plate top electrodes (tin lead for D-1 and D-2, copper foil for D-3) and the readout electrodes were held at a bias of 0~V.
All bias voltages were supplied by an ORTEC~710 power supply.

\paragraph{Radiation source and trigger detector}
An $^{241}$Am source (Eckert \& Ziegler, catalog number AM1A2100U), emitting alpha particles with a mean energy of 4.6~MeV, was mounted on a brass collimator with a length of 80~mm and positioned 95~mm from the center of the hole array.
The collimator features a 1~mm exit aperture.
A PIN photodiode with a sensitive area of 3.2~mm~$\times$~3.2~mm, located at a distance of 15~mm from the center of the dielectric plate, served as the trigger detector and registered the primary alpha particles at a rate of approximately 2.21~Hz.

\paragraph{Low-Pressure Gas System}
The gas system providing continuous gas flow to the low-pressure chamber consisted of a 248A flow control valve (MKS Inc., Andover, MA, USA), an MKS Type~250 pressure/flow controller, an MKS Type~626 Baratron manometer, and a manual precision outflow valve connected to a vacuum pump.
Measurements were performed in propane gas at room temperature $(24\pm1)^\circ$C at a propane gas pressure of 1~mbar.

\paragraph{Data Acquisition}
The nanodosimeter signals were amplified using an ORTEC~9301 fast charge-sensitive preamplifier.
The trigger detector signals were amplified with a CSPA10-1B charge-sensitive preamplifier (FAST ComTec Communication Technology GmbH, 82041 Oberhaching, Germany).
Both preamplifier outputs were connected to the input channels of a Picoscope~3206D digital oscilloscope (Pico Technology Ltd., UK) via 50~$\Omega$ terminators.

\paragraph{Data Processing}\label{sec_data_processing}
The recorded waveforms were first smoothed using a Savitzky–Golay filter, implemented in the \texttt{scipy.signal.savgol\_filter} function from the SciPy \citep{2020SciPy-NMeth} library, to reduce high-frequency noise while preserving the characteristic pulse shape.
Pulse detection was subsequently performed on the filtered signals using \texttt{scipy.signal.find\_peaks} algorithm.

\begin{figure}[htbp]
    \centering
    \includegraphics[width=\linewidth]{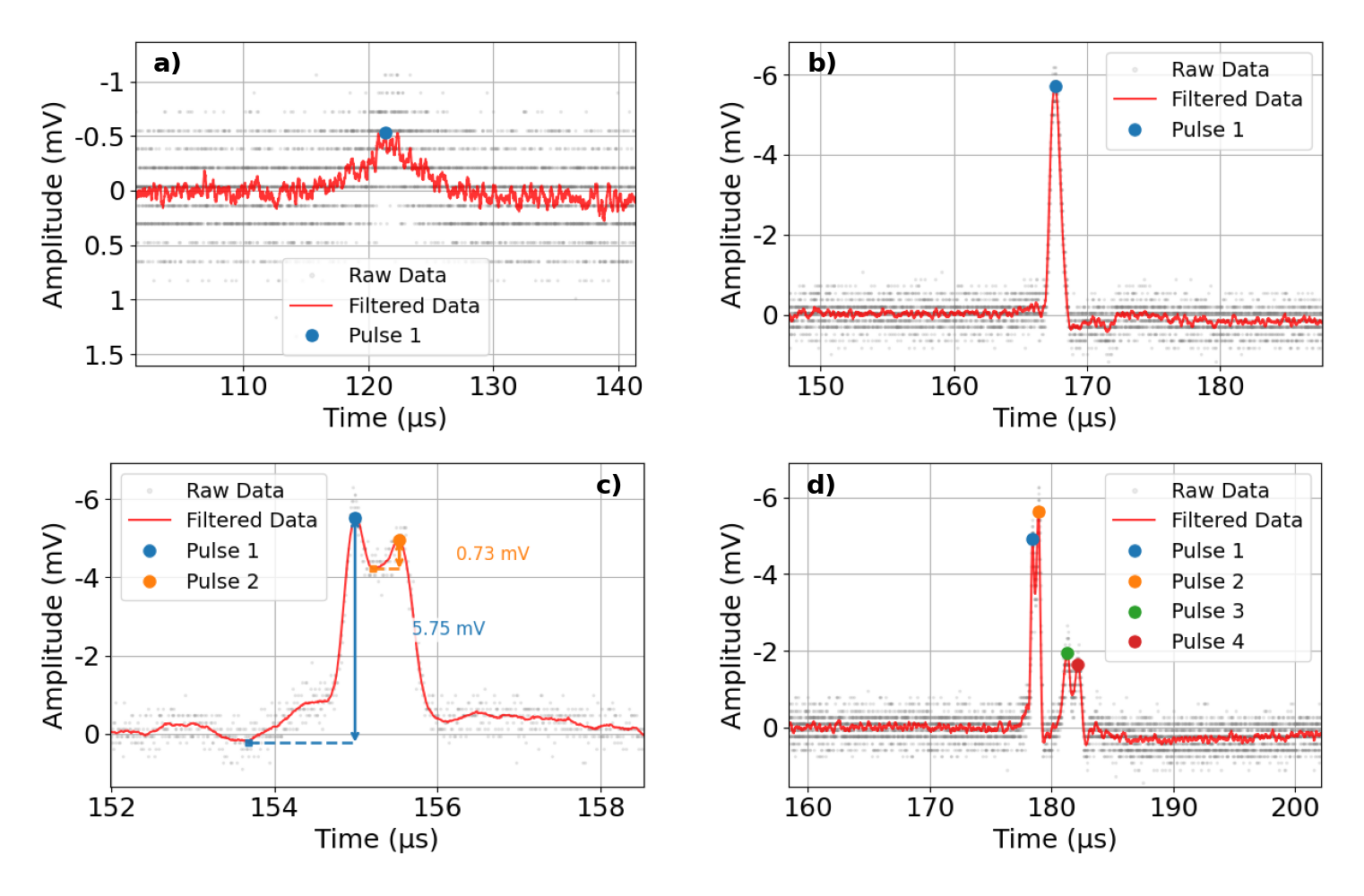}
    \caption{Examples of recorded waveforms and detected pulses for D-1 at a cathode potential of $-800\,\rm V$ (a and b), D-3 at $-800\,\rm V$, with peak prominences indicated by colored arrows (c), and D-3 at $-900\,\rm V$ (d).}
    \label{fig_waveforms}
\end{figure}

For each measurement, the pulse detection parameters were optimized to reliably identify individual pulses while minimizing false detections caused by noise fluctuations.
Fig.~\ref{fig_waveforms} shows examples of waveforms together with the identified pulses.
The standard pulse detection paramteters for the negative pulses included a minimum amplitude threshold of $-0.3~\mathrm{mV}$ for D-1 and D-2 and $-0.4~\mathrm{mV}$ for D-3, a minimum peak prominence of $0.45~\mathrm{mV}$ for D-1 and D-2 and $0.55~\mathrm{mV}$ for D-3, and a minimum and maximum width of $0.16~\text{\textmu s}$ and $14.4~\text{\textmu s}$, respectively.
The peak prominence quantifies the vertical distance between the pulse maximum and the highest local minimum connecting the pulse to a neighboring pulse.
Consequently, the peak prominence criterion suppresses detections originating from small baseline fluctuations while retaining well-separated physical pulses.
The peak prominence is indicated by colored arrows in Fig.~\ref{fig_waveforms}c.
The peak width is defined as the distance between the left and right intersections of the signal with a horizontal line at half the peak prominence.
See \citep{scipy_manual} for further information on the pulse detection parameters.

Waveforms with a pulse multiplicity per alpha trigger $\nu_\mathrm{p}$ were defined as signals containing $\nu_\mathrm{p}$ detected pulses, indexed as \(i=1,\ldots,\nu_\mathrm{p}\).
For each identified pulse, the amplitude, arrival time relative to the corresponding trigger signal, and pulse index \(i\) were determined and stored.
The time window following each trigger, during which the signals were recorded, was 500~\textmu s.
The performance of the pulse detection procedure was validated by visually inspecting at least 300 representative waveforms for every measurement with the detected pulses overlaid.

\subsection{Detector Performance Evaluation}\label{sec_efficiency}
\subsubsection{Charging-up study}
The detector's gain stabilization was investigated by recording the time evolution of the pulse amplitudes during the initial phase of detector operation, immediately after opening the collimator shutter.
To ensure comparable conditions for all charging-up measurements, a conditioning procedure was established.
For each detector configuration, defined as a dielectric plate and cathode potential combination (see Sec.~\ref{meth_cathode_electrodes_voltage}), a small number of signals were initially recorded with the shutter open to verify proper detector operation.
Subsequently, the detector was slowly ramped to an inverse bias of $+500~\mathrm{V}$ applied to the cathode and $-15~\mathrm{V}$ to the anode, and held for approximately 2--3 minutes.
All electrode potentials were then set to $0~\mathrm{V}$, and the detector was left under continuous gas flow for approximately 45 minutes.

Following this period, with the shutter closed, the cathode and anode voltages were gradually ramped to their nominal operating values over the course of several minutes.
The detector was then maintained at the desired operating conditions for an additional two minutes before the shutter was opened and the charging-up measurement commenced.
The 30-minute acquisition period was initiated simultaneously with opening the shutter.

After completion of each measurement, all electrode potentials were again set to $0~\mathrm{V}$ and maintained for at least 10 minutes.
The preparation of the subsequent detector configuration then began by reapplying the inverse bias $+500~\mathrm{V}$ to the cathode and $-15~\mathrm{V}$ to the anode, after which the conditioning procedure described above was repeated.

Complete removal of the charge history of the dielectric plates cannot be guaranteed by this procedure.
However, this procedure was performed to demonstrate the plates' pronounced sensitivity to their charge state.

Three measurement runs were performed at cathode potentials of $-600~\mathrm{V}$, $-700~\mathrm{V}$, and $-800~\mathrm{V}$ for plates D-1 and D-2, and at $-700~\mathrm{V}$, $-800~\mathrm{V}$, and $-900~\mathrm{V}$ for D-3.
The pulse amplitudes were binned into $15~\mathrm{s}$ intervals separately for each pulse index $i$.
The resulting binned datasets were subsequently fitted with a single-exponential function.
The time at which the exponential fit decreased to a value of $1/e$ of its initial value was labeled as the characteristic stabilization time $\tau_i$ for each $i$.
A Student's $t$ factor was applied to determine the combined standard uncertainty of $\tau_i$.
The Student's $t$ factor corrects the uncertainty estimate for finite sample sizes, assuming normally distributed independent measurements.

\subsubsection{Dark Counts Measurement}
The detector was operated until a gain-stabilized state was reached, corresponding to a duration exceeding five times the characteristic stabilization time.
This ensured that stable operating conditions were established before data acquisition.
Subsequently, the shutter was closed, and data acquisition was performed while triggering on the nanodosimeter signals with detection thresholds of $-0.3~\mathrm{mV}$ for D-1 and D-2, and $-0.4~\mathrm{mV}$ for D-3, respectively.
For each configuration, dark-count measurements were performed for approximately $5$--$10~\mathrm{min}$ both prior to and immediately following the main measurement runs.

\subsubsection{Signal Characteristics}
For each measurement series, the cathode potential was increased stepwise from lower to higher biases, with each configuration being recorded for $60~\mathrm{min}$.
Consequently, measurements performed at higher cathode potentials correspond to progressively longer cumulative detector operation times.
The cathode was not cleaned or replaced between individual measurement runs.

For each measurement, the pulse detection algorithm was applied five times using five different parameter combinations (see Sec.~\ref{sec_data_processing}).
The standard parameter set was used as the reference, while the peak height and peak prominence were independently varied by $\pm$10~\%, keeping the other parameter fixed at its standard value.

\section{Results and Discussion}
\subsection{Charging-up of the Dielectric} \label{sec_charging_up}
\begin{figure}[htbp]
    \centering
    \includegraphics[width=0.5\linewidth]{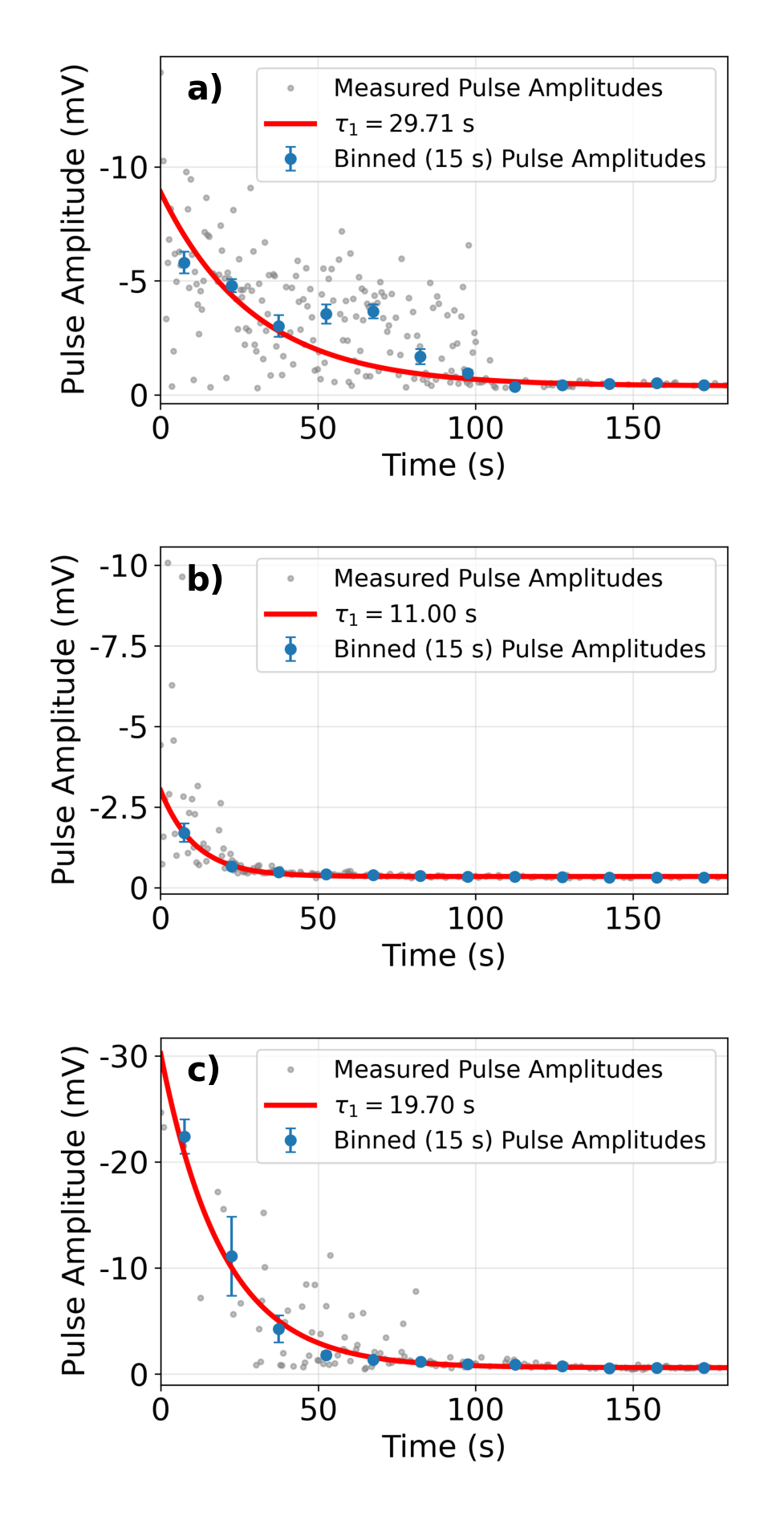}
    \caption{Examples for the measured temporal evolution of the pulse index $i=1$ amplitudes after opening the shutter for a) D-1, b) D-2, and c) D-3, all at a cathode potential of $-700\,\rm V$. The exponential fit is shown as a red line. $\tau_1$ denotes the characteristic stabilization time for the pulses with $i=1$.}
    \label{fig_charging_up}
\end{figure}

Tab.~\ref{tab_characteristic_times} summarizes the mean characteristic stabilization times \(\tau_i\) measured for pulse indices \(i=1,2,3\) for the different investigated detector configurations.
Values reported without uncertainties indicate that the exponential fit was successful for only a single measurement run, owing to insufficient statistics or highly irregular pulse-amplitude evolution.
Blank entries indicate that no reliable exponential fit was obtained for the corresponding pulse index. This becomes increasingly common with increasing pulse index.
Therefore, only pulse indices up to $i=3$ are included in the table.
Fig.~\ref{fig_charging_up}a--c, shows examples of the temporal evolution of $i=1$ pulse amplitudes during the first three minutes after opening the collimator shutter at a cathode potential of \(-700~\rm V\) for the different detector configurations.

\begin{table}[htbp]
    \centering
    \begin{tabular}{l|l|l|l|l}
       Dielectric plate & $U_\mathrm{cath}$ (V) & $\tau_1$ (s) & $\tau_2$ (s) & $\tau_3$ (s) \\ \hline 
       \multirow{3}{*}{D-1}    
         &$-600$& $33\pm 12$ & $21.6$ & -- \\       
         &$-700$& $33\pm 5$ & $31\pm 25$ & -- \\
         &$-800$& $18\pm 8$ & $18\pm 8$ & $24.6$\\ \hline 
         \multirow{3}{*}{D-2}    
         &$-600$& $17.6\pm 1.8$ & -- & -- \\
         &$-700$& $16\pm 6$ & -- & -- \\
         &$-800$& $12\pm 6$ & -- & -- \\\hline
         \multirow{3}{*}{D-3}    
         &$-700$& 19.7 & 14.4 & 15.0 \\
         &$-800$& 9.6 & 6.0 & -- \\
         &$-900$& 3.6 & 7.8 & 16.2 \\
    \end{tabular}
    \caption{Characteristic stabilization times $\tau_i$ for pulse indices $i$ for the stated dielectric plates and cathode potentials.}
    \label{tab_characteristic_times}
\end{table}

For all investigated detector configurations, \(\tau_i\) remained well below one minute despite the relatively low primary particle rate of approximately \(2.21~\rm Hz\).
Overall, the trend was that D-1 exhibited the longest \(\tau_i\) $(18\pm 8)\,\rm s$ to $(33\pm 12)\,\rm s$, whereas D-3 showed the shortest (3.6~s to 19.7~s).
However, the relatively large uncertainties associated with these values should be taken into account when interpreting the results.
For most detector configurations, the  \(\tau_i\) corresponding to different pulse indices were of the same order of magnitude whenever the fitting procedure converged successfully.

For D-1 and D-3, the pulse amplitudes reached a stable plateau following the initial charging-up phase.
In contrast, no observable pulses were registered for D-2 after several  \(\tau_i\), as the pulse amplitudes decreased to the detection threshold of $-0.3\,\rm mV$ within a few minutes.
The observed stabilization behavior is likely associated with a rapid charging-up of the dielectric due to the comparatively large dielectric thicknesses employed in the detector prototypes, namely \(3.175~\rm mm\) for D-1 and D-2 and \(5~\rm mm\) for D-3.

THGEM charging-up was reported to comprise both a short-term process, evolving over minutes, and a long-term process, developing over hours to days \citep{Alexeev2015}.
If a similar separation of charging-up effects applies to the present detector, the measurements presented here cover only the short-term evolution of the process.
Any long-term effects were not investigated.
It should also be noted that most THGEM detectors feature a rim around the holes, whereas the detector used here is rimless.
This difference can have a substantial influence on the charging-up behavior \citep{pitt2018}.

A variety of approaches have been proposed to mitigate charging-up effects in micro-pattern gas detectors, including the application of resistive coatings \citep{song2020} or the use of substrate materials with reduced charge-retention properties \citep{yan2015}.
In principle, similar strategies could also be adapted for compact nanodosimeters.
A detailed investigation of charging-up mitigation techniques was beyond the scope of the present work and is therefore left for future studies.

\subsection{Signal Characteristics}\label{sec_performance_measurements}
In the following, we report on the signal characteristics obtained with D-1 and D-3.
For D-2, no meaningful signals could be obtained, as the detector response rapidly diminished due to charging-up effects.

\subsubsection{Dark counts}
The dark count rates for the investigated detector configurations, measured after stabilization of the detector gain, are summarized in Tab.~\ref{tab_dark_counts}.
The highest dark count rates were measured for (D-1) at a cathode potential of $-900~\rm V$, yielding $(3.3\pm 0.4)~\rm Hz$, and for D-3 at $-1000~\rm V$, yielding $(4.74\pm 0.26)~\rm Hz$.
Since the acquisition window following each trigger was limited to $500$~\textmu s, the probability that a recorded pulse originated from a random dark count rather than from ions generated by the incident alpha particle was negligible.
Assuming Poisson-distributed dark counts, even for the highest measured dark count rate, the expected fraction of such accidental coincidences remained below $0.25~\%$.
Therefore, no dark-count correction was applied during the alpha source measurements.

\begin{table}[htbp]
    \centering
    \begin{tabular}{l|l|l}
       & $U_\mathrm{cath}$ (V) & $R_\mathrm{dark}$ (Hz) \\ \hline 
       \multirow{3}{*}{D-1}    
         &$-700$& $0.023\pm 0.013$   \\       
         &$-800$& $2.43\pm 0.16$  \\
         &$-900$& $3.3\pm 0.4$  \\ \hline 
         \multirow{4}{*}{D-3}    
         &$-700$& $0.142\pm 0.024$ \\
         &$-800$& $2.88\pm 0.12$ \\
         &$-900$& $3.57\pm 0.22$\\
         &$-1000$& $4.74\pm 0.26$  \\
    \end{tabular}
    \caption{Dark count rates $R_\mathrm{dark}$ for the stated detector configurations and cathode potentials $U_\mathrm{cath}$.}
    \label{tab_dark_counts}
\end{table}

\subsubsection{Pulse Amplitude Spectra}\label{res_pulse_amplitudes}
\begin{figure}[htbp]
    \centering
    \hspace*{-0.15\textwidth}%
    \includegraphics[width=1.3\linewidth]{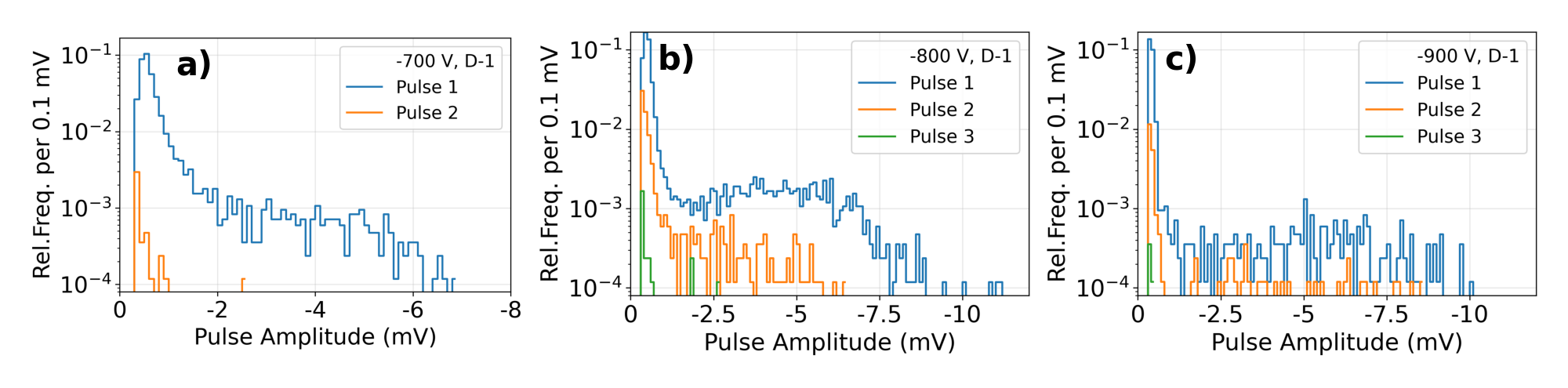}
    \caption{Pulse amplitudes for each pulse index measured with dielectric plate D-1 at cathode potentials of a) $-700\,\rm V$, b) $-800\,\rm V$, and c) $-900\,\rm V$.}
    \label{fig_pulse_amplitude_D1}
\end{figure}

\begin{figure}[htbp]
    \centering
    \includegraphics[width=0.92\linewidth]{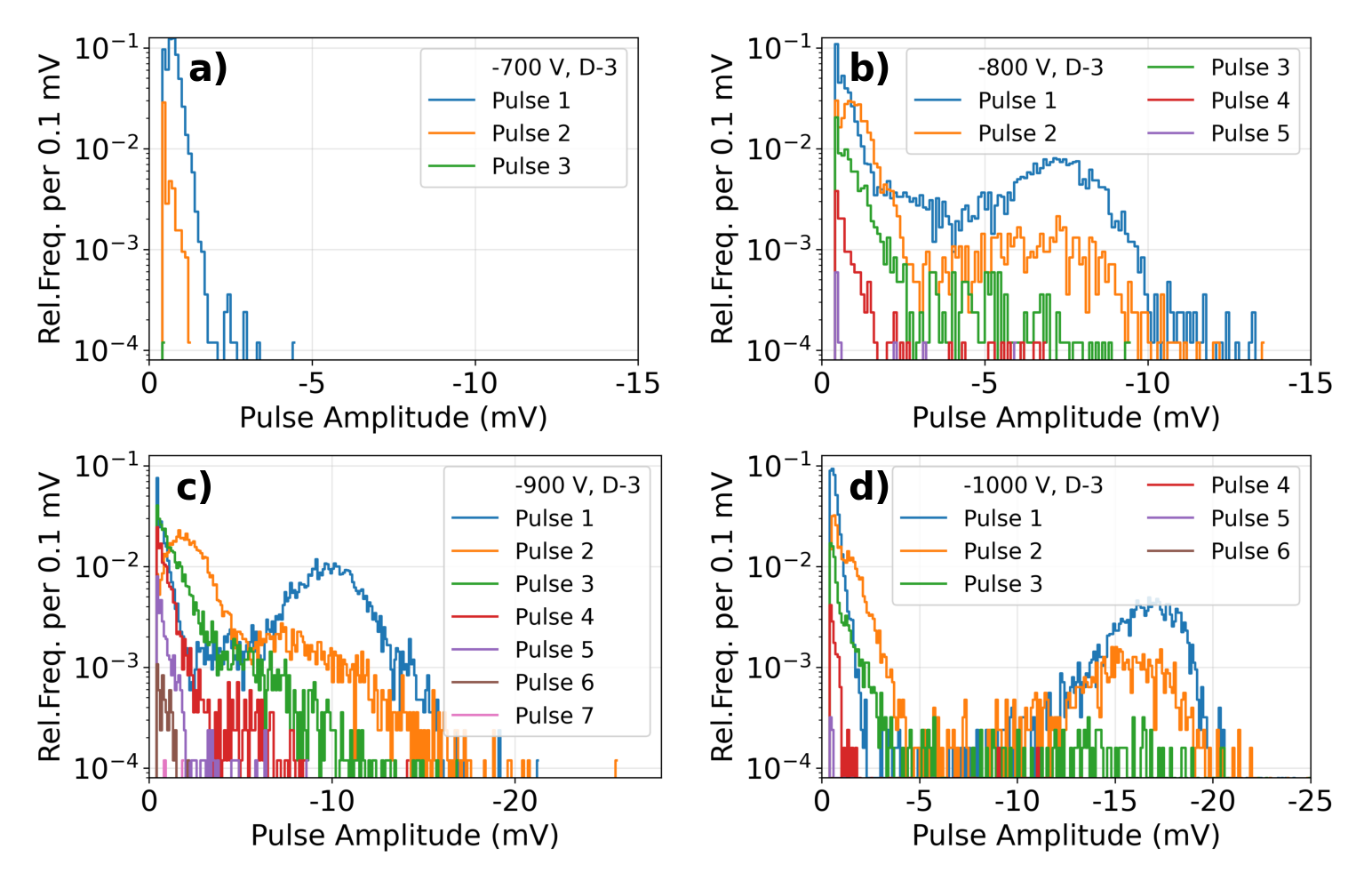}
    \caption{Pulse amplitudes for each pulse index measured with dielectric plate D-3 at cathode potentials of a) $-700\,\rm V$, b) $-800\,\rm V$, c) $-900\,\rm V$, and d) $-1000\,\rm V$.}
    \label{fig_pulse_amplitude_D3}
\end{figure}

Figs.~\ref{fig_pulse_amplitude_D1} and \ref{fig_pulse_amplitude_D3} show the measured pulse amplitude spectra for each pulse index for $\rm C_{1}$ at cathode potentials of $-700~\,\rm V$ to $-900~\,\rm V$, and for $\rm C_{3}$ at $-700~\,\rm V$ to $-1000~\,\rm V$, respectively.
In all cases, the asymmetric pulse amplitude spectra showed peaks close to the detection thresholds of $-0.3\,\rm mV$ for the D-1 measurements and $-0.4\,\rm mV$ for the D-3 measurements.

For D-1 at $-700$~V cathode potential (Fig.~\ref{fig_pulse_amplitude_D1}a), the spectra were dominated by low-amplitude signals of approximately $-1\,\rm mV$ to $-2\,\rm mV$.
At $-800$~V and $-900$~V (Fig.~\ref{fig_pulse_amplitude_D1}b and c), the higher-amplitude tail of the distributions reaching around $-10\,\rm mV$ became more pronounced for pulse indices $i=1,2$.
The infrequently observed pulses with $i=3$ all showed amplitudes close to the D-1 detection threshold of $-0.4\,\rm mV$.

For D-3, the most frequent pulse amplitudes were, likewise to D-1, close to the detection threshold of $-0.4\,\rm mV$ for D-3.
At $-700$~V (Fig.~\ref{fig_pulse_amplitude_D3}a), the spectra are similar to those of D-1, but slightly narrower with a reduced tail toward higher amplitudes, not exceeding $\rm -5\, mV$.
At higher cathode potentials of $-800\,\rm V$ to $-1000\,\rm V$ (Fig.~\ref{fig_pulse_amplitude_D3}b--d), the pulse amplitude spectra became increasingly spread out, accompanied by the emergence of distinct high-amplitude components reaching around $-20\,\rm mV$.
For $-900$~V and $-1000$~V (Fig.~\ref{fig_pulse_amplitude_D3}c and d), $i=1,2$ pulses exhibited broad local maxima around $-9\,\rm mV$ to $-11\,\rm mV$ and $-15\,\rm mV$ to $-17\,\rm mV$, respectively.
Pulses with higher indices did not form a distinct local maximum but instead constituted a long, gradually decaying tail.

Overall, the measurements demonstrated a strong dependence of the pulse amplitudes on the cathode potential.
Increasing the electric field not only enhanced the average pulse amplitude but also resulted in pulse amplitude spectra with an increased spread and more pronounced structure.
This spreading out of the spectra with higher cathode potentials reflects the higher gain at stronger electric fields inside the hole.
The two distinct peaks observed in some spectra may indicate different amplification regimes, e.g., proportional or non-proportional amplification at lower amplitudes and discharge events at higher amplitudes, although the origin remains unclear.
Also, intrinsic statistical fluctuations in the gas gain and variations in the avalanche initiation position within the hole influence the resulting avalanche size and thus the pulse amplitudes.
Both for D-1 and D-3, pulses of indices $i>1$ tended to exhibit slightly reduced amplitudes compared to $i=1$ pulses, while remaining of the same order of magnitude.
This behavior may be influenced by local recharging effects of the resistive cathode, since the available charge for avalanche formation depends on the instantaneous surface charge state.
This state could be modified by preceding avalanches in the same or neighboring holes.
Such effects could potentially be mitigated by individually biasing the holes using separate cathodes.
These could be, e.g., metallic electrodes biased via decoupling or loading resistors.
For pulses occurring close in time, the finite recovery time of the preamplifier may have influenced the measured amplitudes of subsequent pulses.
Of note, separated readout of the individual holes could help identify the origin of the pulses and may clarify the role of correlated avalanche induction and possible secondary peaks, if present.

\subsubsection{Pulse Arrival Time Spectra}
\begin{figure}[htbp]
    \centering
    \hspace*{-0.15\textwidth}%
    \includegraphics[width=1.3\linewidth]{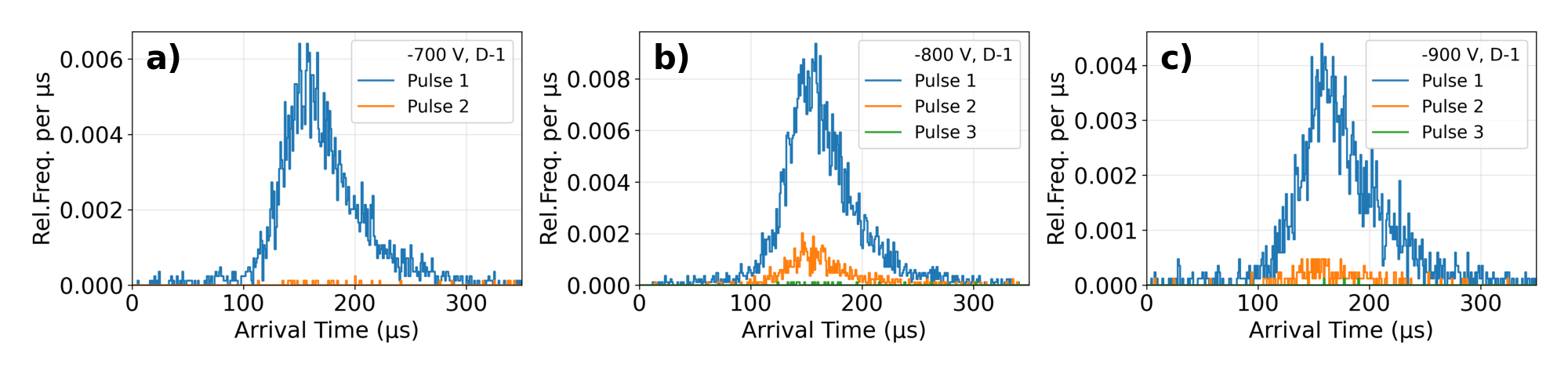}
    \caption{Pulse arrival times for each pulse index measured with dielectric plate D-1 at cathode potentials of a) $-700\,\rm V$, b) $-800\,\rm V$, and c) $-900\,\rm V$.}
    \label{fig_arrival_time_D1}
\end{figure}

\begin{figure}[htbp]
    \centering
    \includegraphics[width=0.92\linewidth]{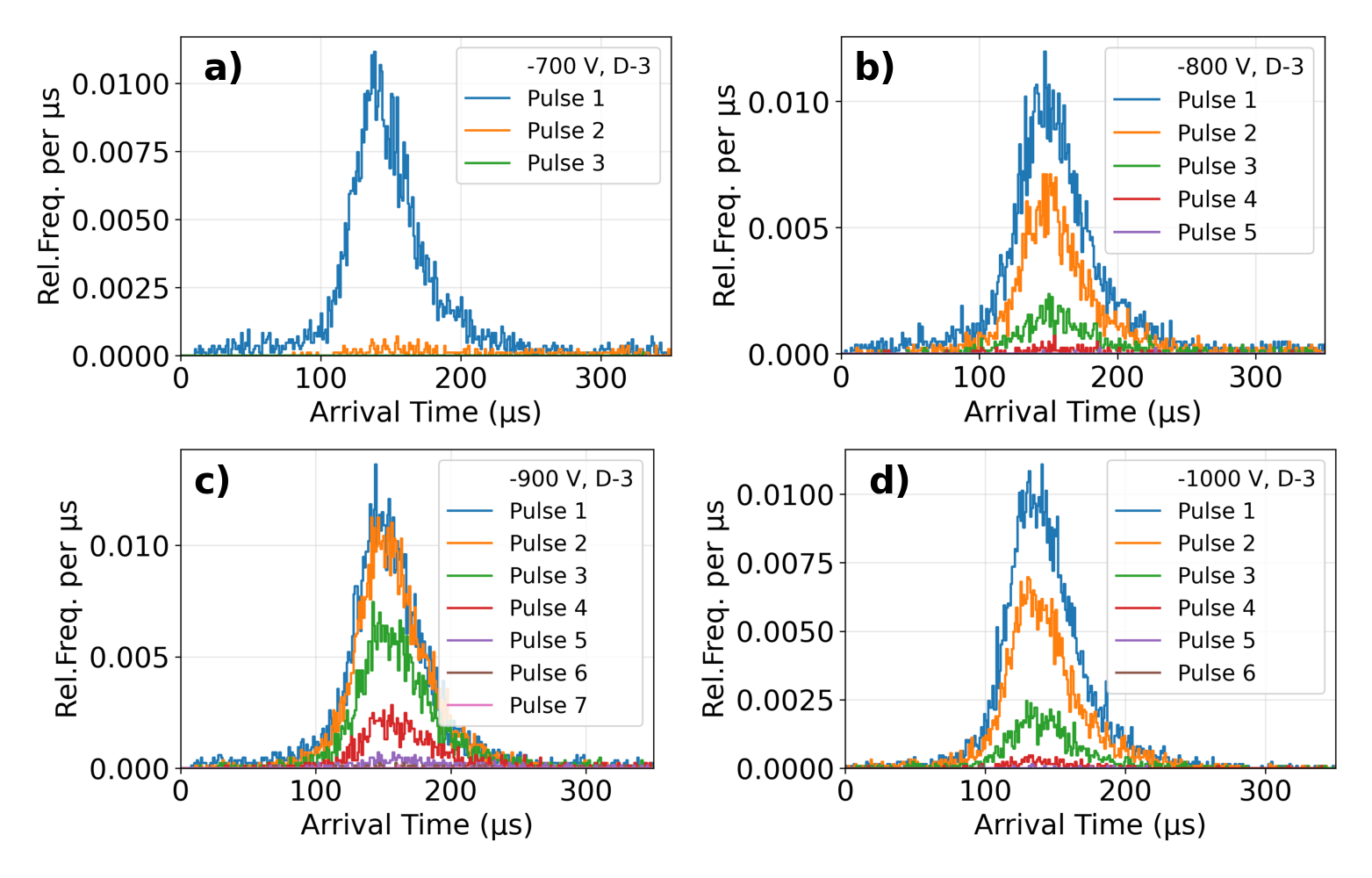}
    \caption{Pulse arrival times for each pulse index measured with dielectric plate D-3 at cathode potentials of a) $-700\,\rm V$, b) $-800\,\rm V$, c) $-900\,\rm V$, and d) $-1000\,\rm V$.}
    \label{fig_arrival_time_D3}
\end{figure}

Figs.~\ref{fig_arrival_time_D1} and \ref{fig_arrival_time_D3} show the pulse arrival time distributions for each pulse index measured with D-1 and D-3, respectively, at the stated cathode potentials.
The pulse arrival time distributions were approximately symmetric but exhibited a slight positive skew, characterized by an extended tail towards longer arrival times.
This tail was more pronounced for D-1 than for D-3, resulting in a correspondingly larger mean pulse arrival time for D-1.
For $i=1$ pulses, mean pulse arrival times for D-1 were around 175~\textmu s to 183~\textmu s with standard deviations typically in the range of around 45~\textmu s to 65~\textmu s at the applied cathode potentials of $-700\,\rm V$ to $-900\,\rm V$.
In contrast, also for $i=1$, D-3 showed systematically earlier mean pulse arrival times at the same cathode potentials.
These were around 155~\textmu s to 166~\textmu s at $-700\,\rm V$ and $-800\,\rm V$ and around 159~\textmu s to 163~\textmu s at $-900\,\rm V$, with a noticeably larger standard deviation of typically 45~\textmu s to 80~\textmu s.
The shift of the mean pulse arrival times toward earlier values for D-3 may be explained by an increased signal formation probability in D-3 compared to D-1 due to the larger thickness of the dielectric, which would lead to earlier signal formation on average.

For pulse indices of $i>1$, both for D-1 and D-3, mean pulse arrival times slightly increased with $i$ by a few \textmu s.
Standard deviations increased significantly, especially for D-3, where standard deviations exceeded 100~\textmu s already at pulse indices of 2–4.
D-1, in comparison, showed a more moderate spread and more stable progression across pulse indices, although statistics became limited with higher $i$, increasing uncertainty.

\begin{figure}[htbp]
    \centering
    \hspace*{-0.15\textwidth}%
    \includegraphics[width=1.3\linewidth]{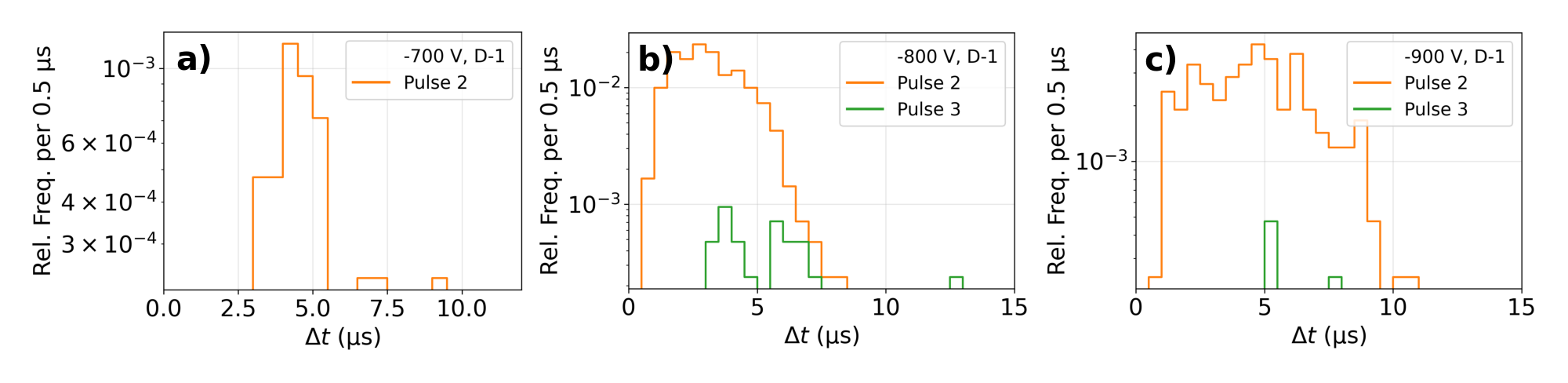}
    \caption{Relative pulse arrival times $\Delta t=t_1-t_i$ measured with dielectric plate D-1 at cathode potentials of a) $-700\,\rm V$, b) $-800\,\rm V$, and c) $-900\,\rm V$.
    $t_1$ is the pulse arrival time of the first pulse of the signal, and $t_i$ is the pulse arrival time of the subsequent pulse with index $i>1$ within the same trigger event.}
    \label{fig_relative_timing_D1}
\end{figure}

\begin{figure}[htbp]
    \centering
    \includegraphics[width=0.92\linewidth]{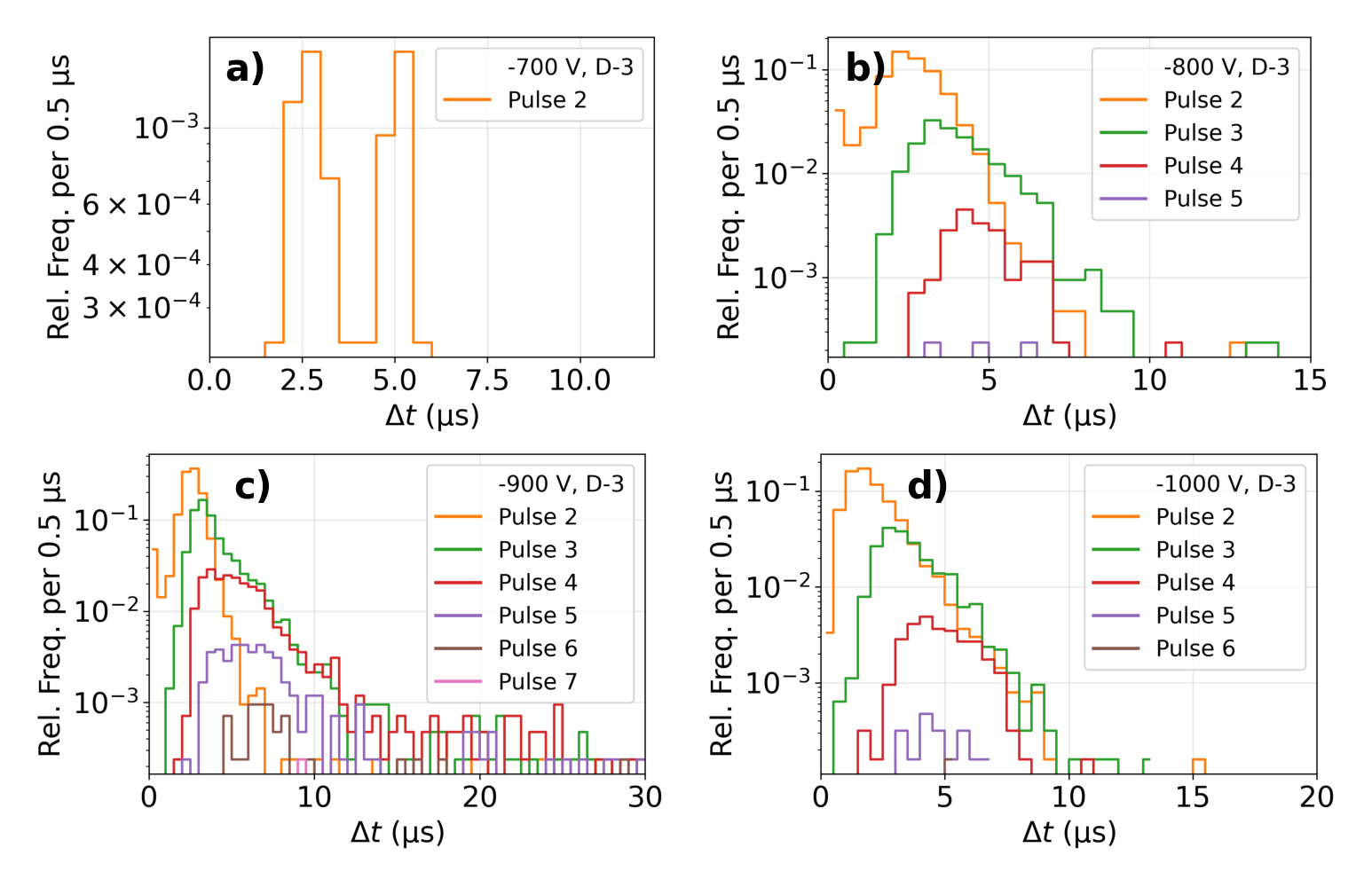}
    \caption{Relative pulse arrival times $\Delta t=t_1-t_i$ measured with dielectric plate D-3 at cathode potentials of a) $-700\,\rm V$, b) $-800\,\rm V$, and c) $-900\,\rm V$, and d) $-1000\,\rm V$.
    $t_1$ is the pulse arrival time of the first pulse of the signal, and $t_i$ is the pulse arrival time of the subsequent pulse with index $i>1$ within the same trigger event.}
    \label{fig_relative_timing_D3}
\end{figure}

Figs.~\ref{fig_relative_timing_D1} and \ref{fig_relative_timing_D3} show the relative pulse arrival times $\Delta t_i$ of pulses with index ${i>1}$ relative to the pulse with $i=1$ within the same trigger event.
Most pulses with $i>1$ occured within 15~\textmu s after the first pulse for both D-1 and D-3.
Only occasionally, longer relative pulse arrival times were observed at any point after the first pulse within the 500~\textmu s time window.
The plots for D-3 further show that the slopes of the probability densities, at least for pulse indices with sufficient statistics, were approximately identical on the right-hand side toward longer time delays.

In these relative pulse arrival time plots, the influence of the collimation of the alpha particle beam on the relative pulse arrival times is largely eliminated, as all pulses were induced by ions originating from the same ionization cluster.
The remaining influence of the collimation is limited to variations in the actual ion drift path length, which may lead to slightly different diffusion of the ions from the ionization cluster to the holes.
Although the setup presented in \citep{merza2026} employed a single-hole compact nanodosimeter with somewhat different design characteristics of the dielectric plate, the ion arrival time spectra simulated in that study may be cautiously compared with the measured pulse arrival time spectra obtained with the present detector.  
This is justified by the identical gas type and pressure, source collimation, anode, and ion drift distance used in both studies.  
For 1~mbar propane, the simulated ion arrival time spectra in \citep{merza2026} exhibited a standard deviation $\sigma$ of approximately 10~\textmu s.  
The vast majority ($>95\,\%$) of the ions arrived within $2\sigma \approx 20$~\textmu s.
Consistent with this result, nearly all pulses, typically more than 90~\%, in the present measurements were observed within a time window of 15~\textmu s.
However, the maxima of the relative pulse arrival time distributions appear fairly regular, which may be related to local recharging and recovery effects, or even the correlated avalanche induction of neighboring holes.

The observation of similar slopes on the right-hand side of the D-3 plots at $-800\,\rm V$ to $-1000\,\rm V$ is physically reasonable, since ions originating from the same ionization cluster and arriving in bunches share a common upper temporal boundary in their arrival times.
Thus, this is a strong indication of the collection of ions in individual holes and independent signal induction. 

\subsubsection{Pulse Multiplicity per Alpha Trigger}\label{sec_pulse_counts}
\begin{figure}[htbp]
    \centering
    \includegraphics[width=\linewidth]{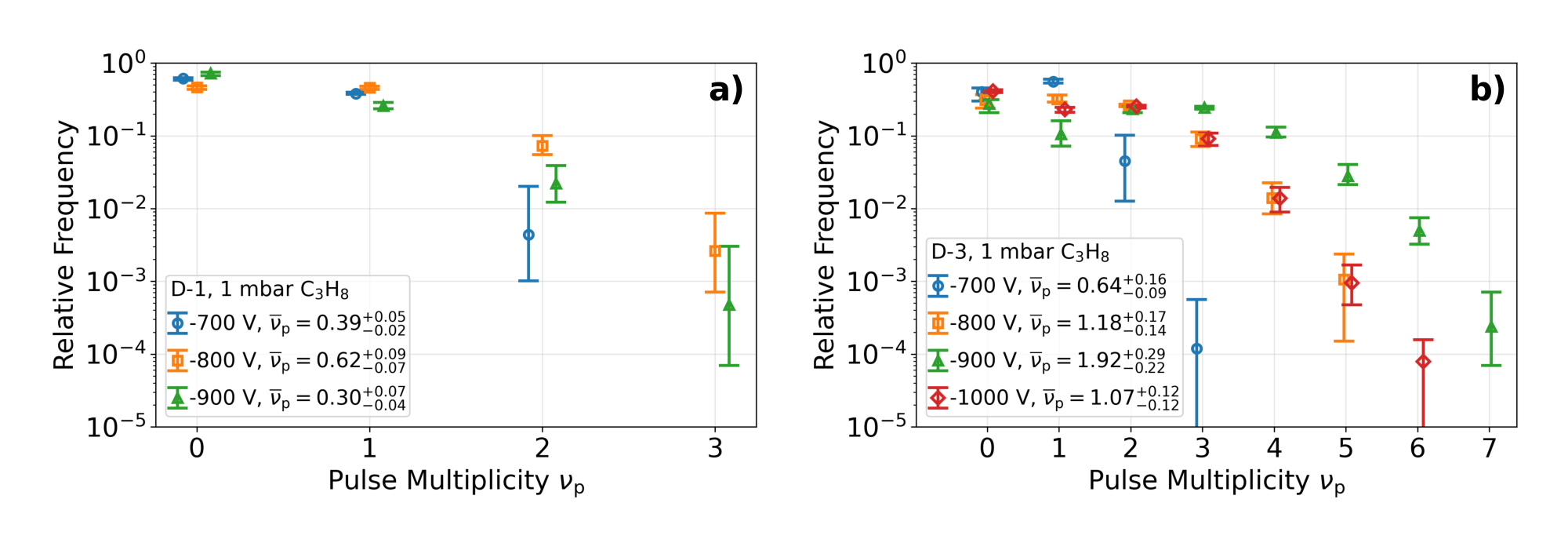}
    \caption{Relative frequency distribution of the pulse multiplicity per trigger $\nu_\mathrm{p}$ measured with a) D-1 and b) D-3 at the stated cathode potentials. Data points were artificially spread within $\pm 0.08$ around the pulse multiplicity tick positions to improve visibility. The lower uncertainty bars of all data points extending beyond the lower y-axis limit reached zero. }
    \label{fig_pulse_counts}
\end{figure}

Fig.~\ref{fig_pulse_counts}a and b show the relative frequency distributions of $\nu_\mathrm{p}$, the pulse multiplicities per alpha trigger, for configurations D-1 and D-3, respectively, including triggers with $\nu_\mathrm{p} = 0$.
The corresponding mean pulse multiplicities per alpha trigger, $\overline{\nu}_\mathrm{p}$, are listed in the legend.
The uncertainties of the relative frequencies of the pulse multiplicities per trigger were estimated by considering both the statistical uncertainty of the results obtained with the standard parameters and the variations in the relative frequencies caused by the parameter variations.
For each upper and lower uncertainty bound, the more conservative value between the statistical standard uncertainty and the deviations resulting from the parameter variations was selected.
Consequently, asymmetrical uncertainty intervals were obtained for the relative frequencies of the pulse indices in most cases.

For D-1 (Fig.~\ref{fig_pulse_counts}a), $\nu_\mathrm{p}$ ranged from 0 to 2 at a cathode potential of $-700\,\rm V$ and from 0 to 3 at cathode potentials of $-800\,\rm V$ and $-900\,\rm V$.
Multiplicities of 0 and 1 dominated for all cathode potentials, followed by a rapid decrease at higher pulse multiplicities.
The highest mean pulse multiplicity per alpha trigger $\overline \nu_\mathrm{p}$ of $\overline \nu_\mathrm{p} = 0.62^{+0.09}_{-0.07}$ was observed at a cathode potential of $-800\,\rm V$.

For D-3, higher $\nu_\mathrm{p}$ than for D-1, reaching up to 7, and more complex relative frequency distributions were observed.
At $-700$~V, the distribution was dominated by 0 and 1 pulse signals.
At $-800$~V, the spectrum shifted further toward higher $\nu_\mathrm{p}$, extending to $\nu_\mathrm{p}=5$.
At $-900$~V, the distribution shifted toward higher $\nu_\mathrm{p}$ but exhibited a non-monotonic shape.
First, $\nu_\mathrm{p}$ decreased from 0 to 1, increased to a local maximum at 3, and decreased again thereafter.
At this cathode potential, the highest mean pulse multiplicity per trigger of $\overline \nu_\mathrm{p} = 1.92^{+0.29}_{-0.22}$ was observed.
At $-1000$~V, the spectrum remained shifted toward higher $\nu_\mathrm{p}$, extending to 6, while showing a similar non-monotonic shape with a local maximum at 2.

Overall, D-3 exhibited higher $\nu_\mathrm{p}$ and higher $\overline \nu_\mathrm{p}$ compared to D-1.
This observation is consistent with a previous study \citep{vasi2016}, which showed that the signal formation probability increases with the thickness of the dielectric structure.
The dielectric plate thickness was 5~mm for D-3 compared with 3.175~mm for D-1.
However, the different dielectric materials (FR-4 for D-1 and acrylic for D-3) and their production method may also contribute to the observed differences in the spectra.

It is not clear why, for both D-1 and D-3, $\overline \nu_\mathrm{p}$ decreased at the highest cathode potentials of $-900\,\rm V$ and $-1000\,\rm V$ compared to the next-lower cathode potentials, at which the highest values of $\overline \nu_\mathrm{p}$ were observed.
One possible explanation is that the larger avalanches produced at higher cathode potentials led to more pronounced cathode discharging, thereby reducing the probability of subsequent pulse generation in neighboring holes.

The origin of the observed non-monotonic pulse multiplicity distributions for D-3 is not yet understood.
Several factors may contribute to this behavior, including charging-up effects of the dielectric, which could require a subsequent relaxation (charging-down) period, as well as dead-time-related effects.

Of note, correlated secondary pulse formation was reported for the FIRE-V2 detector \citep{kempf20252}, which features a functionally very similar working principle.
However, the dielectric plate used in \citep{kempf20252} was considerably thicker (10~mm) than those presented in this work.
These secondary pulses occurred within 0.5~\textmu s to 1.5~\textmu s after the initial pulse and exhibited reduced amplitudes.
Such secondary pulses may also occur in the present setup and are included in the recorded pulse multiplicity if they satisfy the applied pulse detection criteria.
However, due to the readout of several holes by the same electrode, two or more pulses originating from the same hole cannot be distinguished from independently generated pulses in different holes.

Other effects, such as inter-hole coupling via UV photons or secondary electrons generated during an avalanche, may contribute to correlated avalanche induction of neighboring holes.
A dedicated study is required to verify and quantify these effects, if present.

Based on the results and discussion above, the measured pulse multiplicities per trigger cannot yet be referred to as ionization cluster sizes, as the remaining uncertainties and open questions regarding the detector response have not yet been resolved.
The expected mean ionization cluster size for 4.6~MeV alpha particles, which were used in our experiments, was simulated to be on the order of several tens for a single-hole detector with an SV that was similar to or smaller than in our previous study \citep{merza2026}.
In our study, the observed number of pulses is considerably lower than the expected mean ionization cluster size in the SV.
One reason for the apparently low ion detection efficiency is a long dead time.
The observation of a relatively low mean number of pulses per trigger $\overline \nu_\mathrm{p}$ is consistent with the previous study \citep{merza2026} performed under comparable experimental conditions, where the probability of signal formation per collected ion was less than ten percent.

Improvements for future prototypes include additional electrode layers embedded within the dielectric plate, which could optimize ion transport and signal formation, while also reducing the effective dead time by enhancing charge removal following avalanches.
Also, independently powered holes with individual cathodes for each hole, as mentioned in Sec.~\ref{res_pulse_amplitudes},  may further reduce dead time and mitigate inter-hole correlated avalanches.
Planned measurements of IISEE yields from the resistive glass cathode may allow identification of the processes responsible for the observed pulse multiplicity distributions.

\subsection{Measurement Repeatability}\label{sec_repeatability_aging}
\begin{figure}[htbp]
    \centering
    \includegraphics[width=\linewidth]{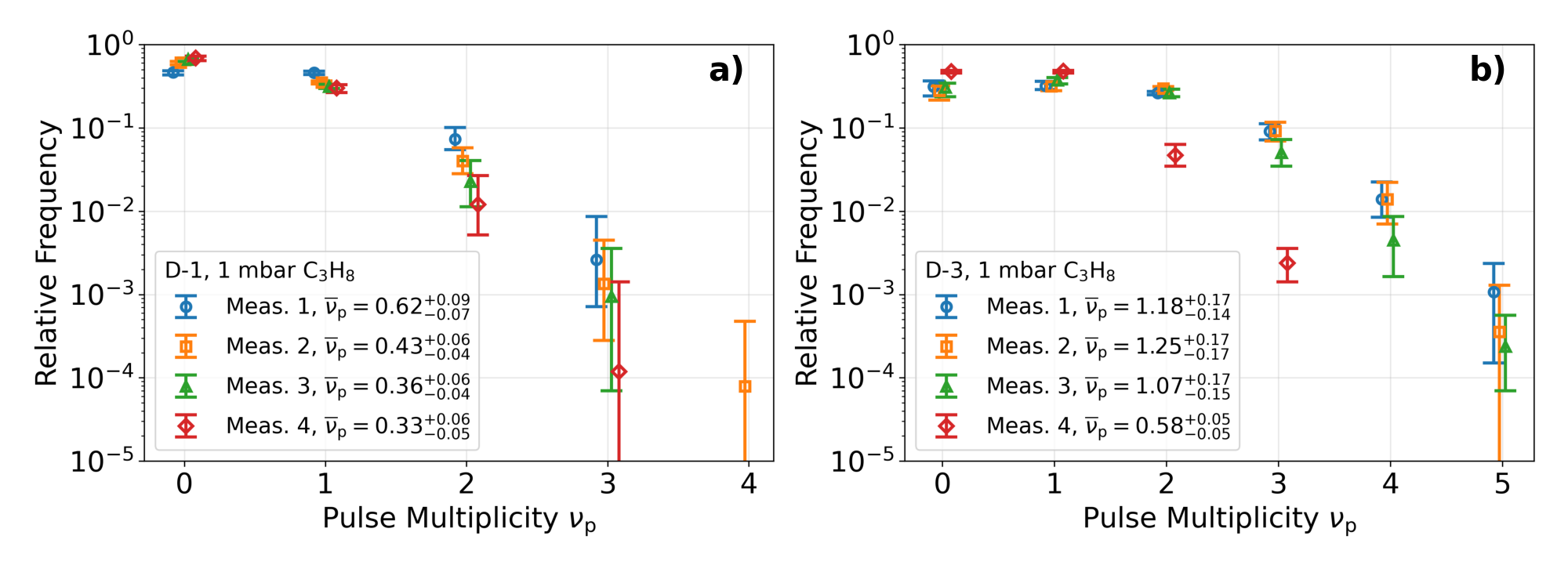}
    \caption{Relative frequency distributions of the pulse multiplicity per trigger for repeated measurements with dielectric plates D-1 (a) and D-3 (b), respectively, at a cathode potential of $-800~\rm V$.
    Data points were artificially spread within $\pm 0.08$ around the pulse multiplicity tick positions to improve visibility.
    The lower uncertainty bars of all data points extending beyond the lower y-axis limit reached zero.
    }
    \label{fig_repeatability_pulse_counts}
\end{figure}

The initial detector performance measurements described in the previous section were repeated three times on consecutive days with at least 14 hours between runs.
All measurements, including the original measurements presented in \ref{sec_performance_measurements}, were performed within two weeks.
Fig.~\ref{fig_repeatability_pulse_counts} shows the repeated relative frequency distributions of the pulse multiplicity per trigger $\nu_\mathrm{p}^{(j)}$ for D-1 and D-3 for the cathode potential of $-800~\rm V$.
The repeated measurements of $\nu_\mathrm{p}^{(j)}$ are labeled from 1 to 4, where 'Measurement 1' is the original measurement.

In general, there was a decrease in $\nu_\mathrm{p}^{(j)} \geq 1$ for D-1 and $\nu_\mathrm{p}^{(j)} \geq 2$ for D-3 over time.
For D-1, the relative frequency of $\nu_\mathrm{p}^{(j)}=0$ and, for D-3, the relative frequencies of $\nu_\mathrm{p}^{(j)}=0,1$ increased, respectively.
The decrease in relative frequency was much more pronounced for $\nu_\mathrm{p}^{(j)} \geq 3$, while the variation over time also increased with multiplicity.
On the other hand, the frequency of $\nu_\mathrm{p}^{(j)}=0$ and, in the case of D-3, also for $\nu_\mathrm{p}^{(j)}=1$, increased over time, which can be explained by the overall decrease in pulse detection.

For D-1, the variations in the relative frequencies of observed pulse multiplicities $\nu_\mathrm{p}^{(j)}$ from Measurement $j=1,...,4$ remained within one order of magnitude for $\nu_\mathrm{p}^{(j)} \leq2$.
Over time, the relative frequency of $\nu_\mathrm{p}^{(j)}=0$ increased, while it decreased for $\nu_\mathrm{p}^{(j)} \geq 1$.
For D-3, the variations among the first three measurements remained comparatively small relative to those observed for D-1.
Measurement 4, however, exhibited a markedly different relative frequency distribution, with substantially lower relative frequencies for $\nu_\mathrm{p}^{(4)} \geq 2$.
The reason for this outlier remains unclear and warrants further investigation.
Although Measurement 2 shows a marginally higher $\overline  \nu_\mathrm{p}^{(2)}$ compared to Measurement 1, the overall trend remained decreasing, in agreement with the behaviour observed for D-1.
The corresponding pulse amplitude and pulse arrival time spectra for the same measurements remained largely unchanged compared to those shown in Sec.~\ref{sec_performance_measurements} (not shown here).

\begin{figure}[htbp]
    \centering
    \includegraphics[width=\linewidth]{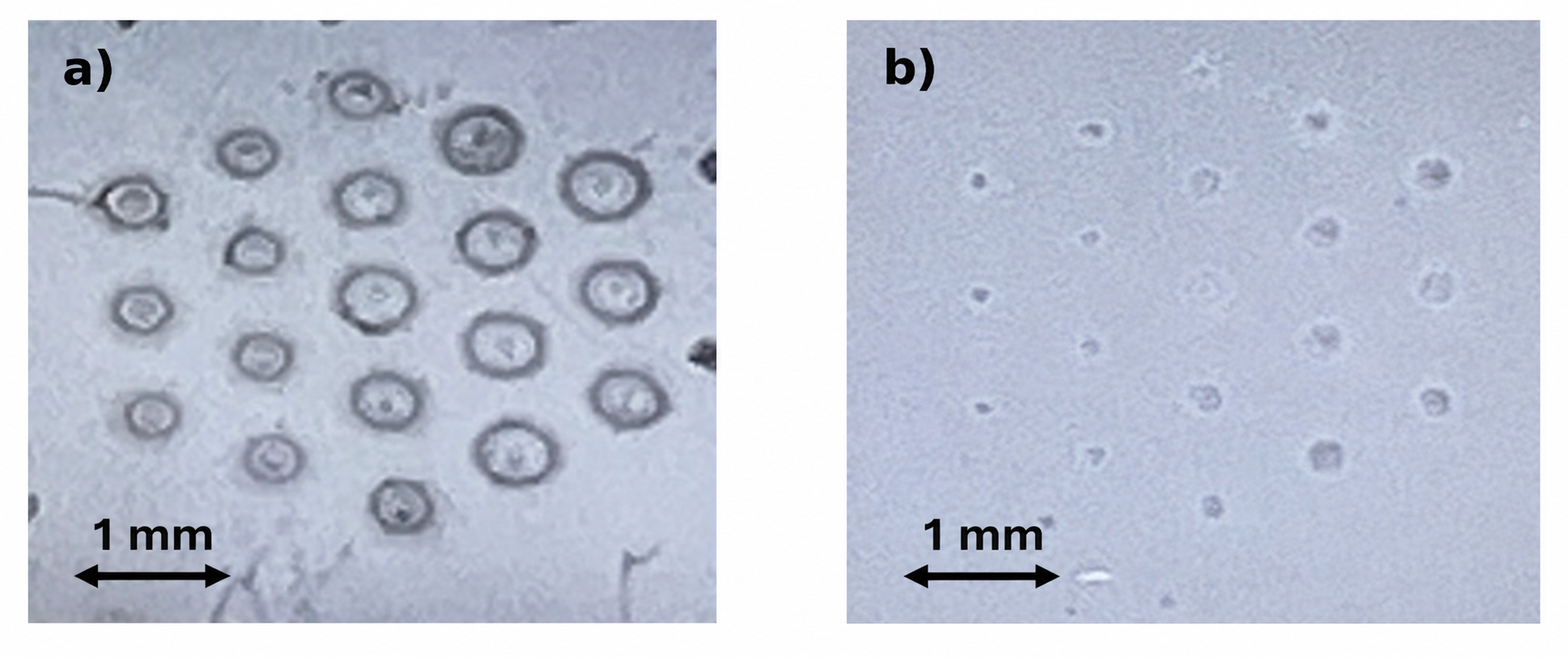}
    \caption{a) Deposits on one of the cathode plates in the areas where it was exposed to the avalanches. b) The same glass cathode after polishing and cleaning it with isopropanol and acetone.}
    \label{fig_depositions}
\end{figure}

After the measurements, the cathode plates were removed from the dielectric plates and visually inspected.
Deposits were observed on the glass surfaces in the regions exposed to the avalanches (see Fig.~\ref{fig_depositions}).
Hydrocarbon gases such as propane are susceptible to polymerization and the formation of carbonaceous deposits during the avalanche, which can explain the observed deposits \citep{capeans2003,sauli2016}.
They have been observed before \citep{FIRE}, and are most likely responsible for the observed aging.
Changes in the electron-emission properties on the glass surface due to deposition may affect the IISEE yield, which may contribute to signal formation \citep{merza2026}.
Using a non-hydrocarbon working gas, such as nitrogen, may reduce detector aging.
Another potential source that may add to detector aging is outgassing from the silicone sealant used to maintain the low-pressure environment at the cathode-dielectric interface \citep{capeans2003}.
Alternative vacuum‑qualified sealants with reduced outgassing characteristics are available and could be implemented in future iterations to minimize outgassing.
We observed the avalanche-induced films forming beneath the holes could be largely removed by sequential cleaning with isopropanol and acetone, but a central spot remained.
This suggests that avalanche‑ and ion-induced modifications extended into the bulk of the glass rather than being confined to the surface.
Such modifications may alter the local electric field, resistivity, and electron-emission properties of the glass and could therefore contribute to long-term performance degradation.

A long-term gain evolution, similar to the one described in \citep{Alexeev2015} attributed to charge mobility within the dielectric, represents another potential aspect of detector stability and may contribute to the observed degradation of detector performance over repeated measurements.
However, this is suspected to be a minor contribution and was not investigated in the present work.

Overall, the detector operated under the present conditions shows only moderate repeatability.
In particular, it is sensitive to the preceding exposure time, which therefore has to be taken into account when comparing measurements.
Consequently, it becomes clear that appropriate conditioning and recovery procedures are required to restore a well-defined initial state and to ensure repeatable measurement results.

\section{Summary and Conclusions}
In this work, a novel nanodosimetric prototype was developed, and its performance was experimentally evaluated using different cathode potentials and three dielectric plates featuring multiple holes.
To the authors’ knowledge, this is the first compact nanodosimeter employing grouped multi-hole amplification structures to extract ions from a single nanometer-equivalent SV.
As such, it represents a step toward the development of a miniaturized nanodosimeter capable of measuring the ICSD.

Two of the dielectric plates, D-1 and D-3, reached a stable operating state after an initial irradiation period, during which pronounced charging-up effects resulted in a rapid decrease followed by stabilization in the pulse amplitudes over a few minutes.
Multiple signal pulses per incident primary particle were observed for all investigated configurations.
The results are consistent with the interpretation that some of the ions produced in the SV during a trigger event are independently registered in the multiple holes.
However, the possibility that an avalanche in one hole induces an avalanche in a neighboring hole cannot be excluded at present.
Further studies are required to quantitatively determine the true ion‑registration pulses induced by independent ions in multiple holes and to distinguish them from secondary pulses generated within the same hole and avalanche‑induced signals in neighboring holes.

While the obtained results indicate that the detector performance is limited, they provide important guidance for the next-generation design.
Approaches for improving the detector design include individual biasing and readout of the holes, as well as layered amplification structures designed to optimize avalanche induction and reduce dead time.
In addition, optimization of the working gas could mitigate the observed detector aging.

The presented nanodosimetric prototype demonstrates promising results and identifies clear pathways for further detector development toward accurate ICSD measurements and practical application.
Compact nanodosimeters could become valuable tools in medical physics and radiation protection, allowing the validation of nanodosimetry-based treatment planning approaches and improved characterization of radiation quality.
Beyond nanodosimetry, this technology may also enable applications in track-structure imaging, space dosimetry, and particle physics.

\section{Author statement}
\textbf{Victor Merza:} Writing - Original Draft, Conceptualization, Methodology.
\textbf{Aleksandr Bancer:} Writing - Review \& Editing, Resources.
\textbf{Vladimir Bashkirov:}
Conceptualization, Methodology.
\textbf{Ana Belchior:} Supervision.
\textbf{Beata Brzozowska:} Writing - Review \& Editing.
\textbf{Piotr Gasik:} Writing - Review \& Editing, Methodology.
\textbf{Jaroslaw Grzyb:} Resources.
\textbf{Khaled Katmeh:} Writing - Review \& Editing.
\textbf{Marcin Pietrzak:} Writing - Review \& Editing.
\textbf{Antoni Ruciński:} Supervision, Project administration.
\textbf{Reinhard Schulte:} Writing - Review \& Editing, Supervision.

\section{Data statement}
All data can be provided by the corresponding author upon request.

\section{Acknowledgements}
This work was partially funded by the Fundação para a Ciência e a Tecnologia (FCT) through the research grants PRT/BD/153748/2021 and PRT/BD/151544/2021.
Further, financial support was received from the National Science Centre, Poland, under the grant number UMO-024/06/Y/ST2/00196.
The authors acknowledge the support of Ingo Deppner (GSI), who kindly provided the low-resistivity glass used in this work.

\bibliographystyle{elsarticle-harv}
\bibliography{paper.bib}

@article{rucinski,
doi = {10.1088/1361-6560/ac35f1},
url = {https://dx.doi.org/10.1088/1361-6560/ac35f1},
year = {2021},
month = {dec},
publisher = {IOP Publishing},
volume = {66},
number = {24},
pages = {24TR01},
author = {Antoni Rucinski and Anna Biernacka and Reinhard Schulte},
title = {Applications of nanodosimetry in particle therapy planning and beyond},
journal = {Physics in Medicine \& Biology},
}

@article{Faddegon,
doi = {10.1088/1361-6560/acea16},
url = {https://dx.doi.org/10.1088/1361-6560/acea16},
year = {2023},
month = {aug},
publisher = {IOP Publishing},
volume = {68},
number = {17},
pages = {175013},
author = {Bruce Faddegon and Eleanor A Blakely and Lucas Burigo and Yair Censor and Ivana Dokic and Naoki Dom\'inguez Kondo and Ramon Ortiz and Jos\'e Ramos M\'endez and Antoni Rucinski and Keith Schubert and Niklas Wahl and Reinhard Schulte},
title = {Ionization detail parameters and cluster dose: a mathematical model for selection of nanodosimetric quantities for use in treatment planning in charged particle radiotherapy},
journal = {Physics in Medicine \& Biology}
}

@article{garty2002,
title = {The performance of a novel ion-counting nanodosimeter},
journal = {Nuclear Instruments and Methods in Physics Research Section A: Accelerators, Spectrometers, Detectors and Associated Equipment},
volume = {492},
number = {1},
pages = {212-235},
year = {2002},
issn = {0168-9002},
doi = {https://doi.org/10.1016/S0168-9002(02)01278-0},
url = {https://www.sciencedirect.com/science/article/pii/S0168900202012780},
author = {G Garty and S Shchemelinin and A Breskin and R Chechik and G Assaf and I Orion and V Bashkirov and R Schulte and B Grosswendt}
}

@article{hilgers2015,
title = {Secondary ionisations in a wall-less ion-counting nanodosimeter: quantitative analysis and the effect on the comparison of measured and simulated track structure parameters in nanometric volumes},
journal = {The European Physical Journal D},
volume = {69},
number = {239},
year = {2015},
doi = {10.1140/epjd/e2015-60176-6},
url = {https://doi.org/10.1140/epjd/e2015-60176-6},
author = {G Hilgers and M U Bug and E Gargioni and H Rabus}
}

@article{hilgers2019,
doi = {10.1088/1748-0221/14/07/P07012},
url = {https://dx.doi.org/10.1088/1748-0221/14/07/P07012},
year = {2019},
month = {jul},
publisher = {},
volume = {14},
number = {07},
pages = {P07012},
author = {G. Hilgers and H. Rabus},
title = {Reducing the background of secondary ions in an ion-counting nanodosimeter},
journal = {Journal of Instrumentation}
}

@article{hilgers2022,
title = {Characterisation of the PTB ion counter nanodosimeter's target volume and its equivalent size in terms of liquid H2O},
journal = {Radiation Physics and Chemistry},
volume = {191},
pages = {109862},
year = {2022},
issn = {0969-806X},
doi = {https://doi.org/10.1016/j.radphyschem.2021.109862},
url = {https://www.sciencedirect.com/science/article/pii/S0969806X21005120},
author = {G Hilgers and T Braunroth and H Rabus}
}

@article{denardo2002,
title = {Ionization-cluster distributions of α-particles in nanometric volumes of propane: measurement and calculation},
journal = {Radiation and Environmental Biophysics},
volume = {41},
pages = {235-256},
year = {2002},
issn = {1432-2099},
doi = {10.1007/s00411-002-0171-6},
url = {https://doi.org/10.1007/s00411-002-0171-6},
author = {L De Nardo and P Colautti and V Conte and W Baek and B Grosswendt and G Tornielli}
}

@article{pszona2000,
title = {A new method for measuring ion clusters produced by charged particles in nanometre track sections of DNA size},
journal = {Nuclear Instruments and Methods in Physics Research Section A: Accelerators, Spectrometers, Detectors and Associated Equipment},
volume = {447},
number = {3},
pages = {601-607},
year = {2000},
issn = {0168-9002},
doi = {https://doi.org/10.1016/S0168-9002(99)01191-2},
url = {https://www.sciencedirect.com/science/article/pii/S0168900299011912},
author = {S Pszona and J Kula and S Marjanska}
}

@article{casiraghi2015,
    author = {Casiraghi, M. and Bashkirov, V. A. and Hurley, R. F. and Schulte, R. W.},
    title = "{Characterisation of a track structure imaging detector}",
    journal = {Radiation Protection Dosimetry},
    volume = {166},
    number = {1-4},
    pages = {223-227},
    year = {2015},
    month = {04},
    issn = {0144-8420},
    doi = {10.1093/rpd/ncv139},
    url = {https://doi.org/10.1093/rpd/ncv139},
    eprint = {https://academic.oup.com/rpd/article-pdf/166/1-4/223/4565335/ncv139.pdf},
}

@article{FIRE,
title = {FIRE: A compact nanodosimeter detector based on ion amplification in gas},
journal = {Nuclear Instruments and Methods in Physics Research Section A: Accelerators, Spectrometers, Detectors and Associated Equipment},
volume = {999},
pages = {165116},
year = {2021},
issn = {0168-9002},
doi = {https://doi.org/10.1016/j.nima.2021.165116},
url = {https://www.sciencedirect.com/science/article/pii/S0168900221001005},
author = {Fabiano Vasi and Irina Kempf and J\"urgen Besserer and Uwe Schneider}
}

@article{grosswendt1,
    author = {Grosswendt, B.},
    title = "{Recent advances of nanodosimetry}",
    journal = {Radiation Protection Dosimetry},
    volume = {110},
    number = {1-4},
    pages = {789-799},
    year = {2004},
    month = {08},
    issn = {0144-8420},
    doi = {10.1093/rpd/nch171},
    url = {https://doi.org/10.1093/rpd/nch171},
    eprint = {https://academic.oup.com/rpd/article-pdf/110/1-4/789/4529014/nch171.pdf},
}

@article{grosswendt2,
    author = {Grosswendt, B. and De Nardo, L. and Colautti, P. and Pszona, S. and Conte, V. and Tornielli, G.},
    title = "{Experimental equivalent cluster-size distributions in nanometric volumes of liquid water}",
    journal = {Radiation Protection Dosimetry},
    volume = {110},
    number = {1-4},
    pages = {851-857},
    year = {2004},
    month = {08},
    issn = {0144-8420},
    doi = {10.1093/rpd/nch203},
    url = {https://doi.org/10.1093/rpd/nch203},
    eprint = {https://academic.oup.com/rpd/article-pdf/110/1-4/851/4529938/nch203.pdf},
}

@phdthesis{bantsar,
    author = "Bantsar, Aliaksandr",
    title = "{Ionization Cluster Size Distributions Created by Low Energy Electrons and Alpha Particles in Nanometric Track Segment in Gases}",
    eprint = "1207.6893",
    archivePrefix = "arXiv",
    primaryClass = "physics.atom-ph",
    school = "Soltan Inst., Swierk",
    year = "2010"
}

@ARTICLE{bashkirov20091,
  author={Bashkirov, V. and Schulte, R. and Wroe, A. and Sadrozinski, H. and Gargioni, E. and Grosswendt, B.},
  journal={IEEE Transactions on Nuclear Science}, 
  title={Experimental Validation of Track Structure Models}, 
  year={2009},
  volume={56},
  number={5},
  pages={2859-2863},
  doi={10.1109/TNS.2009.2029574}}

@INPROCEEDINGS{bashkirov20092,
  author={Bashkirov, V. A. and Hurley, R. F. and Schulte, R. W.},
  booktitle={2009 IEEE Nuclear Science Symposium Conference Record (NSS/MIC)}, 
  title={A novel detector for 2D ion detection in low-pressure gas and its applications}, 
  year={2009},
  volume={},
  number={},
  pages={694-698},
  doi={10.1109/NSSMIC.2009.5402061}}

@article{casiraghi2014,
	Author = {Casiraghi, Margherita and Bashkirov, Vladimir and Hurley, Ford and Schulte, Reinhard},
	Da = {2014/05/07},
	Doi = {10.1140/epjd/e2014-40841-0},
	Id = {Casiraghi2014},
	Isbn = {1434-6079},
	Journal = {The European Physical Journal D},
	Number = {5},
	Pages = {111},
	Title = {A novel approach to study radiation track structure with nanometer-equivalent resolution},
	Ty = {JOUR},
	Url = {https://doi.org/10.1140/epjd/e2014-40841-0},
	Volume = {68},
	Year = {2014}}

@article{vasi2016,
doi = {10.1088/1748-0221/11/09/C09021},
url = {https://dx.doi.org/10.1088/1748-0221/11/09/C09021},
year = {2016},
month = {sep},
publisher = {},
volume = {11},
number = {09},
pages = {C09021},
author = {F. Vasi and M. Casiraghi and V. Bashkirov and U. Giesen and R.W. Schulte},
title = {Development of a single ion detector for radiation track structure studies},
journal = {Journal of Instrumentation}
}

@article{conte2017,
    author = {Conte, V and Selva, A and Colautti, P and Hilgers, G and Rabus, H and Bantsar, A and Pietrzak, M and Pszona, S},
    title = "{NANODOSIMETRY: TOWARDS A NEW CONCEPT OF RADIATION QUALITY}",
    journal = {Radiation Protection Dosimetry},
    volume = {180},
    number = {1-4},
    pages = {150-156},
    year = {2017},
    month = {09},
    issn = {0144-8420},
    doi = {10.1093/rpd/ncx175},
    url = {https://doi.org/10.1093/rpd/ncx175},
    eprint = {https://academic.oup.com/rpd/article-pdf/180/1-4/150/25409814/ncx175.pdf},
}

@Article{conte2023,
AUTHOR = {Conte, Valeria and Bianchi, Anna and Selva, Anna},
TITLE = {Track Structure of Light Ions: The Link to Radiobiology},
JOURNAL = {International Journal of Molecular Sciences},
VOLUME = {24},
YEAR = {2023},
NUMBER = {6},
ARTICLE-NUMBER = {5826},
URL = {https://www.mdpi.com/1422-0067/24/6/5826},
PubMedID = {36982899},
ISSN = {1422-0067},
DOI = {10.3390/ijms24065826}
}

@article{pietrzak2018,
  author    = {Pietrzak, M. and Pszona, S. and Bantsar, A.},
  title     = {Measurements of Spatial Correlations of Ionisation Clusters in the Track of Carbon Ions-First Results},
  journal   = {Radiation Protection Dosimetry},
  volume    = {180},
  number    = {1-4},
  pages     = {162--167},
  year      = {2018},
  doi       = {10.1093/rpd/ncy079},
  url       = {https://doi.org/10.1093/rpd/ncy079}
}

@article{bancer2020,
title = {Particle track structure measurements from 0.5 to 18 nm in nitrogen using the Jet Counter nanodosemeter},
journal = {Radiation Physics and Chemistry},
volume = {172},
pages = {108805},
year = {2020},
issn = {0969-806X},
doi = {https://doi.org/10.1016/j.radphyschem.2020.108805},
url = {https://www.sciencedirect.com/science/article/pii/S0969806X19308126},
author = {Aleksandr Bancer and Marcin Pietrzak and Monika Mietelska}
}

@article{merza2025,
title = {Garfield++ and Geant4-DNA simulation of a compact THGEM-based nanodosimeter},
journal = {Nuclear Instruments and Methods in Physics Research Section A: Accelerators, Spectrometers, Detectors and Associated Equipment},
volume = {1080},
pages = {170729},
year = {2025},
issn = {0168-9002},
doi = {https://doi.org/10.1016/j.nima.2025.170729},
url = {https://www.sciencedirect.com/science/article/pii/S0168900225005303},
author = {Victor Merza and Aleksandr Bancer and Ana Belchior and Beata Brzozowska and João F. Canhoto and Khaled Katmeh and Marcin Pietrzak and Antoni Ruciński and Reinhard Schulte}
}

@article{kempf20252,
title = {Development and characterization of a compact nanodosimetric detector},
journal = {Nuclear Instruments and Methods in Physics Research Section A: Accelerators, Spectrometers, Detectors and Associated Equipment},
volume = {1075},
pages = {170337},
year = {2025},
issn = {0168-9002},
doi = {https://doi.org/10.1016/j.nima.2025.170337},
url = {https://www.sciencedirect.com/science/article/pii/S016890022500138X},
author = {Irina Kempf and Tamara Melina Hoffmann and Jürgen Besserer and Uwe Schneider}
}

@article{wang2010,
title = {Development of multi-gap resistive plate chambers with low-resistive silicate glass electrodes for operation at high particle fluxes and large transported charges},
journal = {Nuclear Instruments and Methods in Physics Research Section A: Accelerators, Spectrometers, Detectors and Associated Equipment},
volume = {621},
number = {1},
pages = {151-156},
year = {2010},
issn = {0168-9002},
doi = {https://doi.org/10.1016/j.nima.2010.04.056},
url = {https://www.sciencedirect.com/science/article/pii/S0168900210009058},
author = {Jingbo Wang and Yi Wang and Xianglei Zhu and Weicheng Ding and Yuanjing Li and Jianping Cheng and Nobert Herrmann and Ingo Deppner and Yapeng Zhang and P. Loizeau and P. Senger and D. Gonzalez-Diaz}
}

@article{wang2019,
doi = {10.1088/1748-0221/14/06/C06015},
url = {https://doi.org/10.1088/1748-0221/14/06/C06015},
year = {2019},
month = {jun},
publisher = {},
volume = {14},
number = {06},
pages = {C06015},
author = {Wang, Y. and Zhang, Q. and Han, D. and Wang, F. and Yu, Y. and Lyu, P. and Li, Y.},
title = {Status of technology of MRPC time of flight system},
journal = {Journal of Instrumentation}
}

@misc{merza2026,
      title={Experimental and Monte Carlo Simulation Studies to Investigate the Working Principle of Compact Nanodosimeters}, 
      author={Victor Merza and Aleksandr Bancer and Vladimir Bashkirov and Ana Belchior and Beata Brzozowska and João F. Canhoto and Piotr Gasik and Jaroslaw Grzyb and Khaled Katmeh and Marcin Pietrzak and Antoni Ruciński and Reinhard Schulte},
      year={2026},
      eprint={2512.11126},
      archivePrefix={arXiv},
      primaryClass={physics.ins-det},
      url={https://arxiv.org/abs/2512.11126}, 
}

@article{heylen1975,
author = {A. E. D. Heylen},
title = {Ionization coefficients and sparking voltages from methane to butane},
journal = {International Journal of Electronics},
volume = {39},
number = {6},
pages = {653--660},
year = {1975},
publisher = {Taylor \& Francis},
doi = {10.1080/00207217508920532},
URL = {https://doi.org/10.1080/00207217508920532},
eprint = {https://doi.org/10.1080/00207217508920532}
}

@article{Alexeev2015,
doi = {10.1088/1748-0221/10/03/P03026},
url = {https://doi.org/10.1088/1748-0221/10/03/P03026},
year = {2015},
month = {mar},
publisher = {},
volume = {10},
number = {03},
pages = {P03026},
author = {Alexeev, M. and Birsa, R. and Bradamante, F. and Bressan, A. and Büchele, M. and Chiosso, M. and Ciliberti, P. and Torre, S. Dalla and Dasgupta, S. and Denisov, O. and Duic, V. and Finger, M. and Jr, M. Finger and Fischer, H. and Gobbo, B. and Gregori, M. and Herrmann, F. and Königsmann, K. and Levorato, S. and Maggiora, A. and Makke, N. and Martin, A. and Menon, G. and Novakova, K. and Novy, J. and Panzieri, D. and Pereira, F.A. and Santos, C.A. and Sbrizzai, G. and Schiavon, P. and Schopferer, S. and Slunecka, M. and Sozzi, F. and Steiger, L. and Sulc, M. and Takekawa, S. and Tessarotto, F. and Veloso, J.F.C.A.},
title = {The gain in Thick GEM multipliers and its time-evolution},
journal = {Journal of Instrumentation}
}

@article{song2020,
title = {Production and properties of a charging-up “Free” THGEM with DLC coating},
journal = {Nuclear Instruments and Methods in Physics Research Section A: Accelerators, Spectrometers, Detectors and Associated Equipment},
volume = {966},
pages = {163868},
year = {2020},
issn = {0168-9002},
doi = {https://doi.org/10.1016/j.nima.2020.163868},
url = {https://www.sciencedirect.com/science/article/pii/S0168900220303685},
author = {Guofeng Song and Ming Shao and Lunlin Shang and Yi Zhou and You Lv and Xu Wang and Jianbei Liu and Zhiyong Zhang}
}

@article{pitt2018,
doi = {10.1088/1748-0221/13/03/P03009},
url = {https://doi.org/10.1088/1748-0221/13/03/P03009},
year = {2018},
month = {mar},
publisher = {},
volume = {13},
number = {03},
pages = {P03009},
author = {Pitt, M. and Correia, P.M.M. and Bressler, S. and Coimbra, A.E.C. and Renous, D. Shaked and Azevedo, C.D.R. and Veloso, J.F.C.A. and Breskin, A.},
title = {Measurements of charging-up processes in THGEM-based particle detectors},
journal = {Journal of Instrumentation}
}

@article{yan2015,
doi = {10.1088/1674-1137/39/6/066001},
url = {https://doi.org/10.1088/1674-1137/39/6/066001},
year = {2015},
month = {jun},
publisher = {Chinese Physical Society and the Institute of High Energy Physics of the Chinese Academy of Sciences and the Institute of Modern Physics of the Chinese Academy of Sciences and IOP Publishing},
volume = {39},
number = {6},
pages = {066001},
author = {Yan, Jia-Qing and Xie, Yu-Guang and Hu, Tao and Lu, Jun-Guang and Zhou, Li and Qu, Guo-Pu and Cai, Xiao and Niu, Shun-Li and Chen, Hai-Tao},
title = {Simulation and performance study of ceramic THGEM*},
journal = {Chinese Physics C}}

@ARTICLE{2020SciPy-NMeth,
  author  = {Virtanen, Pauli and Gommers, Ralf and Oliphant, Travis E. and
            Haberland, Matt and Reddy, Tyler and Cournapeau, David and
            Burovski, Evgeni and Peterson, Pearu and Weckesser, Warren and
            Bright, Jonathan and {van der Walt}, St{\'e}fan J. and
            Brett, Matthew and Wilson, Joshua and Millman, K. Jarrod and
            Mayorov, Nikolay and Nelson, Andrew R. J. and Jones, Eric and
            Kern, Robert and Larson, Eric and Carey, C J and
            Polat, {\.I}lhan and Feng, Yu and Moore, Eric W. and
            {VanderPlas}, Jake and Laxalde, Denis and Perktold, Josef and
            Cimrman, Robert and Henriksen, Ian and Quintero, E. A. and
            Harris, Charles R. and Archibald, Anne M. and
            Ribeiro, Ant{\^o}nio H. and Pedregosa, Fabian and
            {van Mulbregt}, Paul and {SciPy 1.0 Contributors}},
  title   = {{{SciPy} 1.0: Fundamental Algorithms for Scientific
            Computing in Python}},
  journal = {Nature Methods},
  year    = {2020},
  volume  = {17},
  pages   = {261--272},
  adsurl  = {https://rdcu.be/b08Wh},
  doi     = {10.1038/s41592-019-0686-2},
}

@misc{scipy_manual,
  author       = {{Sci{P}y Developers}},
  title        = {Sci{P}y {API} {R}eference},
  year         = {2026},
  howpublished = {\url{https://docs.scipy.org/doc/scipy/reference/}},
  note         = {Accessed: 2026-07-15}
}

@article{Ortiz2025,
    doi = {10.1088/1361-6560/ae07a3},
    url = {https://doi.org/10.1088/1361-6560/ae07a3},
    year = {2025},
    month = {sep},
    publisher = {IOP Publishing},
    volume = {70},
    number = {19},
    pages = {195004},
    author = {Ortiz, Ramon and Ramos-Méndez, José and Mao, Jian-Hua and Schulte, Reinhard and Faddegon, Bruce},
    title = {Evaluation of nanodosimetric quantities for ion radiotherapy treatment planning based on the degree of association of survival with cluster dose},
    journal = {Physics in Medicine \& Biology}
}

@article{sauli2016,
title = {The gas electron multiplier (GEM): Operating principles and applications},
journal = {Nuclear Instruments and Methods in Physics Research Section A: Accelerators, Spectrometers, Detectors and Associated Equipment},
volume = {805},
pages = {2-24},
year = {2016},
note = {Special Issue in memory of Glenn F. Knoll},
issn = {0168-9002},
doi = {https://doi.org/10.1016/j.nima.2015.07.060},
url = {https://www.sciencedirect.com/science/article/pii/S0168900215008980},
author = {Fabio Sauli}
}

@article{capeans2003,
title = {Aging and materials: lessons for detectors and gas systems},
journal = {Nuclear Instruments and Methods in Physics Research Section A: Accelerators, Spectrometers, Detectors and Associated Equipment},
volume = {515},
number = {1},
pages = {73-88},
year = {2003},
note = {Proceedings of the International Workshop on Aging Phenomena in Gaseous Detectors},
issn = {0168-9002},
doi = {https://doi.org/10.1016/j.nima.2003.08.134},
url = {https://www.sciencedirect.com/science/article/pii/S0168900203024550},
author = {M. Capeans}
}

\end{document}